\documentclass[sigconf]{acmart}
\AtBeginDocument{%
  \providecommand\BibTeX{{%
    \normalfont B\kern-0.5em{\scshape i\kern-0.25em b}\kern-0.8em\TeX}}}

\usepackage{caption}
\usepackage{subcaption}

\graphicspath{ {figures/} }

\usepackage{color}
\usepackage{soul}    
\usepackage{gensymb} 

\newcommand{\new}[1]{\textcolor{black}{#1}}
\newcommand{\revise}[1]{\textcolor{black}{#1}}
\newcommand{\recheck}[1]{\textcolor{black}{#1}}

\usepackage{enumitem} 

\usepackage{xspace} 

\newcommand{\systemName}{\textsc{Draw2Cut}\xspace}

\copyrightyear{2025}
\acmYear{2025}
\setcopyright{acmlicensed}\acmConference[CHI '25]{CHI Conference on Human Factors in Computing Systems}{April 26-May 1, 2025}{Yokohama, Japan}
\acmBooktitle{CHI Conference on Human Factors in Computing Systems (CHI '25), April 26-May 1, 2025, Yokohama, Japan}
\acmDOI{10.1145/3706598.3714281}
\acmISBN{979-8-4007-1394-1/25/04}

\begin{document}

\title[\systemName: Direct On-Material Annotations for CNC Milling]{\systemName:  Direct On-Material Annotations for CNC Milling}


\author{Xinyue Gui}
\authornote{These authors contributed equally to this work.}
\affiliation{%
  \institution{The University of Tokyo}
  \city{Tokyo}
  \country{Japan}}
\email{xinyueguikwei@gmail.com}

\author{Ding Xia}
\authornotemark[1]
\affiliation{%
  \institution{The University of Tokyo}
  \city{Tokyo}
  \country{Japan}}
\email{dingxia1995@gmail.com}

\author{Wang Gao}
\affiliation{%
  \institution{The University of Tokyo}
  \city{Tokyo}
  \country{Japan}}
\email{gaowang@g.ecc.u-tokyo.ac.jp}

\author{Mustafa Doga Dogan}
\affiliation{%
  \institution{Adobe Research}
  \city{Basel}
  \country{Switzerland}}
\email{doga@adobe.com}

\author{Maria Larsson}
\affiliation{%
  \institution{The University of Tokyo}
  \city{Tokyo}
  \country{Japan}}
\email{ma.ka.larsson@gmail.com}

\author{Takeo Igarashi}
\affiliation{%
  \institution{The University of Tokyo}
  \city{Tokyo}
  \country{Japan}}
\email{takeo@acm.org}



\begin{abstract}
Creating custom artifacts with computer numerical control (CNC) milling machines typically requires mastery of complex computer-aided design (CAD) software. To eliminate this user barrier, we introduced \systemName, a novel system that allows users to design and fabricate artifacts by sketching directly on physical materials. \systemName employs a custom-drawing language to convert user-drawn lines, symbols, and colors into toolpaths, thereby enabling users to express their creative intent intuitively. The key features include real-time alignment between material and virtual toolpaths, a preview interface for validation, and an open-source platform for customization. Through technical evaluations and user studies, we demonstrate that \systemName lowers the entry barrier for personal fabrication, enabling novices to create customized artifacts with precision and ease. Our findings highlight the potential of the system to enhance creativity, engagement, and accessibility in CNC-based woodworking.
\end{abstract}


\begin{teaserfigure} 
  \centering
  \includegraphics[width=0.99\textwidth]{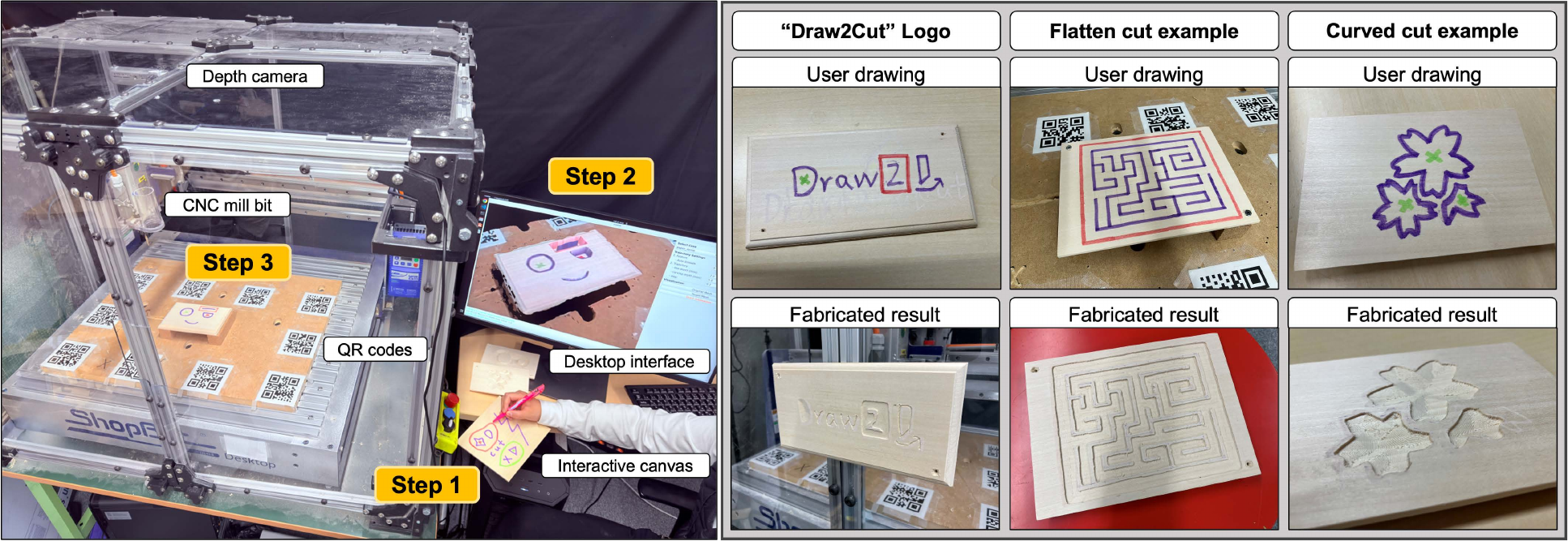}
  \caption{Left: Overview of the three steps for creating artifacts using the \systemName system.
  \textit{Step 1:} Users input their fabrication intent by drawing on the physical material to be cut.
  \textit{Step 2:} Users place the annotated wood in the work space and checks the preview.
  \textit{Step 3:} The CNC machine mills the material automatically, allowing users to choose further processing (e.g., assemble, paint).
  Right: Sample artifacts produced with \systemName: a logo, a maze, and a \textit{sakura} flower.}
  \Description{Left panel: Three steps for creating artifacts using the \systemName system: \textit{(1)} The user inputs their fabrication intent by drawing on the physical material to be cut. \textit{(2)} The user places the annotated wood in the work space and checks the preview. \textit{(3)} CNC machine mills the wood automatically, allowing users the option of further processing (e.g., assemble or paint). Right panel: Sample artifacts in different dimensions produced with \systemName: a designed logo, a wooden maze, and carved \textit{sakura} patterns. There are three modules. The \textit{language module} interprets user-drawn curves and symbols in several colors into machine-readable instructions. The \textit{registration module} aligns the user’s annotations in physical space with virtual space and returns the computing result to the physical world of the same size. The \textit{preview module} provides a cut preview and animation. The user’s drawing pen is erasable. If they are not satisfied with the cut preview result, then they can erase the curve and re-draw it.}
  \label{fig:1}
\end{teaserfigure}

\maketitle


\section{Introduction}  

An experienced fabricator (e.g., carpenter) uses tools and experiential techniques to create a scheme to fabricate artifacts. However, this knowledge needs to be gained from experience because it is not codified \cite{tran_oleary_taxon_2021}. Digital fabrication has attracted considerable attention because of the emergence of \textit{personal fabrication} \cite{gershenfeld_fab_2005, subbaraman_3d_2023, mueller_tutorial_2014}. Numerous computer-aided design (CAD) software have been developed for designing and operating controllable machines for fabrication \cite{bangse_design_2020}. These tools can bridge the gap between complex fabrication and rapid prototyping \cite{magrisso_digital_2018, tran_oleary_taxon_2021}, enabling designers to convert ideas into objects without the need for handcraft knowledge. However, a significant gap remains between complex fabrication and rapid prototyping \cite{twigg-smith_tools_2021, yildirim_digital_2020, landwehr_sydow_modding_2022}, given the limitations identified in the three workflow steps.

The first step involved the measurement and calibration of the materials. Subtractive manufacturing involves using existing materials. Therefore, the size, shape, and relative position of the material in the machine bed before cutting are critical. Typically, CAD users are required to measure or scan materials to extract sufficient data (e.g., length, width, and height) to create a digital canvas as the foundation. Moreover, prior to cutting, CNC users must calibrate the default setting of the machine to align it with the cut target. These tedious tasks are necessary but have little to do with creative design. In this stage, \textbf{working directly in real space} is more intuitive than planning using a CAD interface, because users can quickly obtain a sense of size. Certain projects (e.g., \textit{interactive construction} \cite{mueller_interactive_2012}) can be implemented and \revise {some commercial tools (e.g., \textit{Shaper Trace}\footnote{\url{https://www.shapertools.com/en-us/trace}} and \textit{Glowforge}\footnote{\url{https://shop.glowforge.com/products/glowforge-pro}}) can be operated in real space;} however, these approaches require users to draw on a separate interface at a 1:1 scale rather than directly on the material. Working on a separate interface makes 2D planning effortless, but these are limited to subtractive manufacturing products that have a relative spatial relationship with the existing object (e.g., ~\cite{zheng_joinery_2017}).

The second step is to create a digital blueprint in CAD. The process of producing, modifying, and altering geometries (displayed as vectors) to produce target objects requires the user to master the software. Even experts face challenges when dealing with irregular materials or modifying a blueprint during fabrication \cite{de-los-aires-solis_wood-wood_2023}, because expanding or modifying the work spaces for machines is difficult \cite{peek_cardboard_2017}. Instead of editing objects through an interface, it is simpler for users to draw contours \textit{in physical space} by \textbf{directly drawing on the material} using a pen. Hand-drawing offers users a high degree of freedom to create objects with flexible shapes and sizes. Currently, some projects are focused on a certain product. \textit{Maker’s Mark} \cite{savage_makers_2015} modified a 3D shape for 3D printing using only printouts as primary input. However, this was for additive fabrication; they performed scans to collect processed object spatial information, whereas the subtractive fabrications were required to perform 3D registration to align the virtual space with the real-world coordinates. \textit{CopyCAD} \cite{follmer_copycad_2010} focused on subtractive fabrication by enabling the user to copy and paste real-world objects onto a CNC canvas. However, these methods are limited to two-dimensional (2D) shapes. \new{We explore 3D fabrication methods. Three-axis CNC machines have limited DOF because they can only achieve vertical subtraction. Therefore, we make a trade off between several limitations to investigate limited 3D (extensive x- and y-ranges, limited z-range) subtractive fabrication \cite{rout2022fabrication, 2.5D}.}

The third step is toolpath generation. This step can be completed by clicking and selecting options in CAD after finishing the blueprint design, or it can be customized in G-code, which is the most frequently used CNC and 3D printing programming language. This step requires users to be familiar with the cut material and hardware settings (types of milling bits and feed rate) that correspond to the cutting type (e.g., a ball cutter for smooth transitions between cuts, different angle V-bits for V-carving, an end mill cutter for pocketing, and the option to generate straight or arc slots). Previous studies have indicated that this technique is difficult for inexperienced users. Hirsch et al. \cite{hirsch_nothing_2023} interviewed 13 professional designers and identified difficulties in computer-aided manufacturing (CAM). For example, to perform an actual cut, enabling only CNC movement followed by the user’s drawing curve, as in a sketch-based CNC \cite{jacobs_sketch-based_2017}, was insufficient. The machine should be instructed on \textit{how to cut} (e.g., where to start and how deep to cut). We propose adapting some conventional mark syntax (e.g., cross marks and zigzag marks) from carpenters while making them CNC-readable. This strategy allows us to \textbf{simplify the interactive process of designing toolpaths while maintaining product diversity}.

Motivated by these considerations, we propose a novel approach called \systemName that enables users to directly convert their design intent into a final product. The user draws the wood and places it on a machine bed. The CNC machine then scans and cuts automatically following the user’s drawings of the contours and cutting instruction symbols. In this study, a two-stage process was used. We first defined a design space and subsequently conducted a fast-prototyping route using a CAD-based prototype to determine the technical requirements. Based on the findings, we divided the development process into three modules: 1) A 3D space registration system that aligns the physical working space with the virtual computing space. 2) The hand-drawn content in the tool path mapping language allows users to create machine-readable instructions using three color pens. Handwriting has a high degree of freedom as an input. For the fabrication intent, different users sketched differently because of their drawing habits. Therefore, a precise mapping language was required. 3) After interpreting the user intent, a separate user interface (UI) displays the cutting preview from a perspective. Our code is \textbf{open-source}\footnote{\url{https://github.com/ApisXia/Draw2Cut}} and enables other designers or fabrication artists to produce their one-of-a-kind artifacts using our designed language system.

We present technical evaluation (Section 6) and demonstration (Section 7) to show the feasibility and effectiveness of \systemName. In the technical evaluation, we assess the precise positioning of the cutting, measure the influence of material thickness, and determine the minimum size of the feature. In the demonstrations, we manufacture several items in four parts progressively. We first demonstrate basic advantage of the proposed system, which shows that \systemName is a better option for CAD in certain scenarios. We then demonstrate the effectiveness of our method for traditional woodworking tasks. Third, we run a workshop to demonstrate that it is actually usable by end users. Fourth, we report additional demonstration to show advanced results (e.g., complexity). The contributions of this study are as follows:
\vspace{-0.05cm}
\begin{itemize}[leftmargin=0.5cm]
    \item We propose an interactive system, \systemName, which lowers the entry barrier for novice users engaged in personal fabrication using CNC milling. The system facilitates mapping between the virtual design and the physical material.
    \item We provide a drawing language (e.g., mark syntax) that maps multiple inputs (lines, symbols, and text) with attributes (colors) to precise machine behaviors (a blueprint with a specific toolpath).
    \item We present technical evaluation and four demonstrations: quantitative precision measurements and qualitative evaluation by demonstration, and discussed the scope of the use cases.
\end{itemize}

\section{Related Work} 

\subsection{Lowering Technical Barriers for Personal Fabrication}

The canonical digital fabrication \cite{twigg-smith_tools_2021} comprises several parts. We categorized them into two categories, namely using CAD to design the product in digital form, and using CAM to physically manufacture the output. Typically, CAD is considered an essential component of CAM \cite{what_is_CAM_2021}. Therefore, focusing on computer-aided design (CAD) for fabrication is a popular research topic. Researchers have added numerous functionalities to software such that users can handle it more easily \cite{peng_--fly_2016, peng_d-coil_2015}. Concurrently, designers have developed CAD interfaces to make visualizations more intuitive and comprehensive \cite{vasquez_jubilee_2020, fossdal_interactive_2021, moyer2024throwing}. These studies have focused on software interactions. Another study direction is to remove CAD from the idea-to-fabrication process. Eliminating CAD could decrease the complexity, precision, and reproducibility \cite{peng_roma_2018}. However, user interactions are not confined to the software. After combining mixed reality, multimodal sensing, and other technologies, various interactive environments or approaches can emerge. This phenomenon lowers the technical barrier for people who want to rapidly prototype creative ideas (that is, personal fabrication). Therefore, we focused on this direction.

Another difficulty encountered after removing the CAD is translating the user’s varied inputs into instructions that the processing machine can execute. CAD provides an interface for users to select the depth, cutting methods, and cutting tools, which are then converted into toolpaths (that is, G-code output files). Consequently, previous CAD studies have focused on toolpath generation. For example, researchers have developed an interface capable of displaying several types of data, not just geometry \cite{tran_oleary_improving_2022}, allowing for all digital attributes \cite{peng_d-coil_2015} and designing toolpaths so that they can be altered parametrically during manufacturing \cite{zoran_human-computer_2013}. For challenging designs, such as wooden joints, researchers have developed supporting tools, such as interfaces, which can provide a detailed picture of the joints \cite{larsson_tsugite_2020} or incorporate a parametric joint library to enable rapid exploration \cite{tian_matchsticks_2018}. \textit{SensiCut} incorporates a material-sensing platform that provides a user interface with material information to assist users in determining the appropriate power and speed while avoiding dangerous material cutting \cite{dogan_sensicut_2021}. If we remove the typical CAD (i.e., screen-free) rather than considering the input method, then we should consider other methods to generate precise G-codes for CNC.

\subsection{Input Methods and Interactive Environment}

There are various studies to address the problem of interaction efficiency and alignment accuracy between physical and digital spaces. Previous studies addressed this issue through several approaches. For examples, the multimodal sensing approaches rendered the forms of interactions more diverse. The enriched input types included voice, temperature, pressure, gestures, and touch. \textit{FormFab} reshapes a thermoplastic material using a heat gun \cite{mueller2019formfab}, \textit{Speaker} shapes a wire based on the user’s voice (sensing a simplified sound wave), and \textit{Cutter} incorporates a thin wire heated by electrical resistance to control the cutter’s pushing or pulling \cite{willis_interactive_2010}. To adjust the shape of the polyurethane foam beneath the touchscreen, users input the pattern on the touchscreen. Mueller et al. presented an interactive construction \cite{mueller_interactive_2012} in which the user uses a laser pointer to draw lines on the screen above the target material within a laser cutter and subsequently cuts the material using a CNC. However, in these studies, rather than sketching directly on a physical item, the user was required to use a separate instrument (such as a drawing pad or heat gun).

The second category includes methods based on augmented reality (AR)~\cite{dogan_fabricate_2022, dogan_augmented_2024}. Users physically work on a product using mixed reality (MR) devices as intermediaries to ensure that their instructions are coordinated with the virtual model~\cite{iyer_xr-penter_2025, arslan_realitycraft_2024}. For example, \textit{SPATA} examined integration into virtual design environments using data from two geographical measurements \cite{weichel_spata_2015}. \textit{RoMA} allows customers to expand existing items with an AR controller or create well-proportioned physical artifacts \cite{peng_roma_2018}. AR-based techniques are useful for remote collaboration fabrication \cite{goveia_da_rocha_exquisite_2021, kim_machines_2017, tian_turn-by-wire_2019, tian_adroid_2021} and manufacturing, which demand real-time feedback \cite{takahashi_interactive_2016, albaugh_augmented_2023}. For example, in fabrications that require constant adjustments during the process (e.g., clay and ceramics \cite{frost2024sketchpath}), complex expressions can only be exposed through the embodied process of production \cite{landwehr_sydow_modding_2022, tokac_craft-inspired_2022}. \revise{The commercial tool, \textit{Shaper Origin}\footnote{\url{https://www.shapertools.com/en-us/origin}}, a handheld CNC router, allows a user to cut directly on the material with real-time feedback using an augmented-reality overlay.} MR allows continuous or turn-taking inputs, enabling fabrication and design benefits through feedback loops and creative inputs. Because the material we used (e.g., wood) was relatively stable during the process, we did not achieve such real-time interactivity.

However, as stated in the \textit{RoMA} limitations \cite{peng_roma_2018}, the fabrication quality and precision of the calibration are difficult to manage because of inaccuracies throughout the overlay procedure between the physical and digital models. Another issue comes from the property in subtractive manufacturing. Adding or deforming fabrication may only require aligning the physical input with the virtual environment and executing computing. The user can deform the material based on the calculated virtual results \cite{mueller2019formfab}. A 3D printing machine can directly produce outcomes in an organized form \cite{frost2024sketchpath}. A framework \textit{ModelCraft} \cite{song2009modelcraft} employs a user input approach similar to ours but focuses on physical-to-digital synchronization rather than actual fabrication. However, in subtractive manufacturing, after completing the computation in the virtual space, the system should cut in the real space at the correct point relative to the material to be cut. In other words, the proposed module should perform a secondary calibration to ensure that the virtual model matches the physical space. A project-camera system is required to complete the registration.

For example, \textit{CopyCAD} \cite{follmer_copycad_2010} allows users to copy 2D shapes from real-world objects captured by a camera and project these shapes exactly where the genuine objects are. \textit{Maker’s Mark} \cite{savage_makers_2015} did not incorporate a standard camera, but used 3D scanning to collect spatial information about the processed product. Jacobs et al. developed a sketch-based CNC \cite{jacobs_sketch-based_2017}, which allows CNC movement, followed by a drawing pattern of the same size as the drawn line. This approach projects an augmented object onto different spaces to enhance interaction techniques. For example, \textit{iLamps} investigated several interaction strategies to describe object augmentation using a handheld projector \cite{raskar_ilamps_2006}. \textit{Shader Lamps} directly recreated the appearance of a real item and boosted its visual quality using a projector \cite{raskar_shader_2001}. We applied a similar method to the subtractive manufacturing of wood using a camera to compute the environmental information for CNC cutting. In other words, we aligned our physical workspace with the underlying virtual space while calibrating the computed digital results into physical material.

\subsection{Manufacturing Process Execution}

Some studies have focused on CAM procedures that simplify the programming process, such as creating a creative coding environment \cite{subbaraman_p5_2022}. "Physical-digital programming" first originated for tailoring fabrication workflows for practitioners with varied programming language underpinnings \cite{tran_oleary_physical-digital_2023}. Based on this approach, \textit{Extruder-Turtle} was developed to generate G-code for engineers familiar with the open-source \textit{Turtle Geometry} library \cite{pezutti-dyer_extruder-turtle_2022}.  Researchers developed \textit{StickyLand}, which organizes code in computational notebooks for data scientists such as \textit{Jupyter} notebooks~\cite{wang_stickyland_2022}. Their primary objective was to create a programming environment with which users were already comfortable. However, the target users remain constrained to those with programming skills. 

Researchers have generalized toolpath editing by inventing \textit{Imprimer}, which can detect various documents, including interleaved text and pictures, and is suitable for users from various backgrounds \cite{tran_oleary_imprimer_2023}. This approach is appropriate for patterns that are easily described in literature. However, this approach still struggles with irregular shapes that are difficult to describe verbally. Consequently, sketch-based methods have been devised \cite{bremers_computer_2022, willis_spatial_2010}. \textit{SketchPath} enables a 3D printer to interpret the user’s hand-drawn textures and use them to organize 3D clay printing shapes \cite{frost2024sketchpath}. In addition to additive manufacturing, our project obtained a similar method to perform subtractive fabrication using hand-drawn symbols as the cut-type instructions to generate the toolpath.

\section{Design Criteria}

We propose using handwritten instructions (e.g., lines and symbols) to bridge the gap between virtual modeling and physical fabrication. To develop the design criteria, we first investigated the envisioned interactions to identify possible input and output methods. We subsequently developed a fast prototyping system using an existing tool to identify technical requirements. This section focuses on the following questions.

\begin{itemize}[leftmargin=0.5cm]
\item \textit{How is the proposed workflow feasible and effective? (Sec. 3.1)}
\item \textit{What technical adjustments are needed to ensure a seamless and accurate system? (Sec. 3.2)}
\end{itemize}

\subsection{Envisioned Interaction}

\begin{figure*}[!h]
\includegraphics[width=0.9\textwidth]{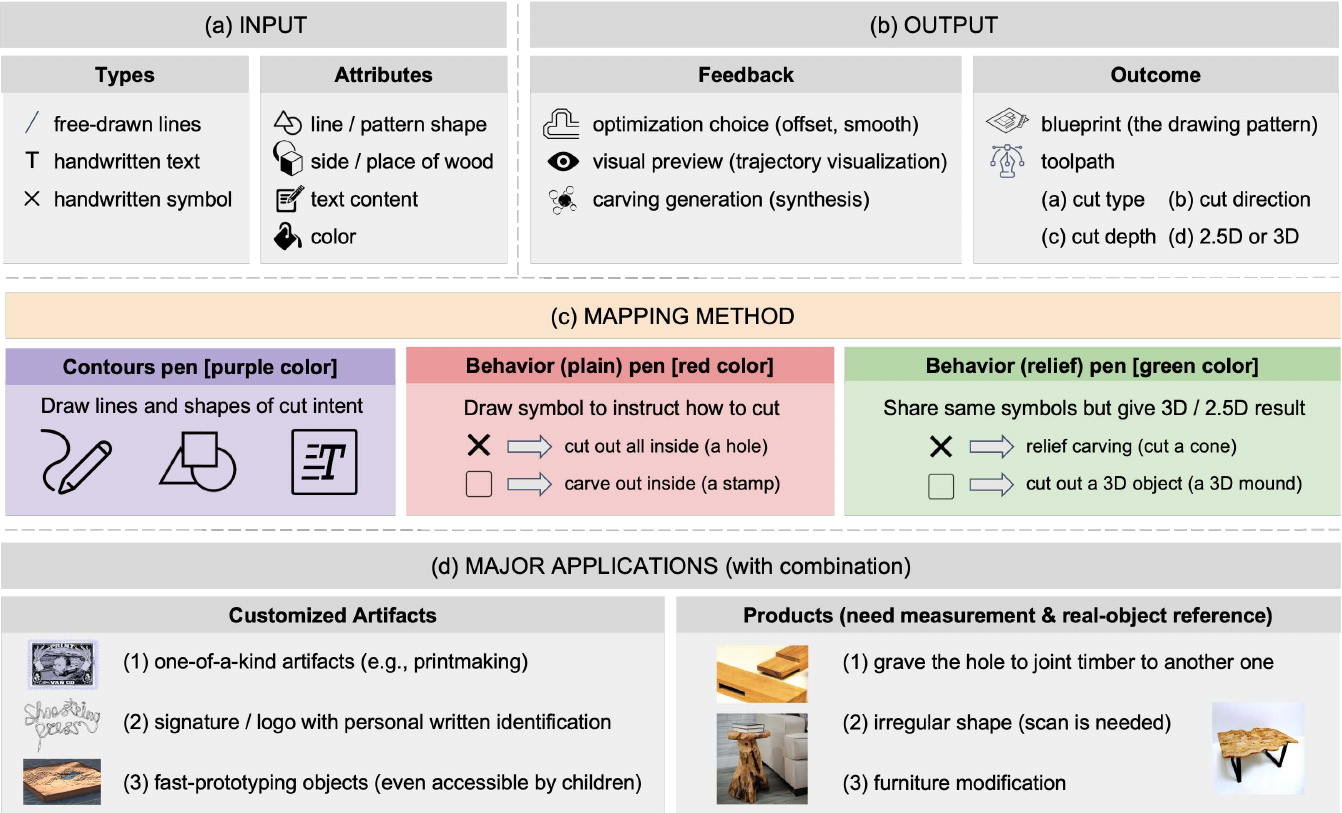}
\caption{Design space for exploring the interaction methods, including four aspects: (a) possible input, (b) necessary output, (c) procedure of how the drawing language is developed, and (d) application scope.}
\Description{Design space for exploring the interaction methods, including four aspects: (a) possible input, (b) necessary output, (c) procedure of how the drawing language is developed, and (d) application scope.}
\label{fig:2}
\end{figure*}

We first propose a design space considering four design-related issues outlined below (Figure~\ref{fig:2}a-d).

\paragraph{What may be an input from a drawing (Figure~\ref{fig:2}, part a)?} 
Within the constraint that all inputs come solely from the user’s hand drawings on the physical material, we analyzed the types and attributes of possible inputs. In terms of type, users can draw straight lines or curves to construct different forms, write various texts, and draw symbols to communicate specific instructions. In terms of attributes, we can use various color pens and choose the relative locations of different marks to convey various cutting intentions.

\paragraph{What type of outputs are desired (Figure~\ref{fig:2}, part b)?}
To investigate the user demand, we surveyed existing CAD software for CNC fabrication, such as \textit{VCarve Pro}\footnote{\url{https://www.vectric.com/support/tutorials/vcarve-pro/?}} and \textit{Aspire}\footnote{\url{https://www.vectric.com/products/aspire/}}, and obtained example demos. In these applications, we divided user actions into three stages. The first action involves creating a blueprint. Users select vectors (e.g., letters, circles) and manipulate them to form an outline. The second action is to generate the toolpaths. Users then select the profile (engraving the outline) or pocket (cutting the area inside) choices to specify how to cut each vector. The third action is to validate the results using a preview interface. CAD offers cut animation and preview of expected outcomes.

\paragraph{How should we create a mapping language to connect the input and output (Figure~\ref{fig:2}, part c)?}

CAD system generates a unique TAP file, which is a G-code file. This indicates that the user’s cut intent, CAD output G-code, and machine movement have a one-to-one relationship. However, when the user input shifted to handwriting, the degree of flexibility increased. Even if the fabrication intent is the same, different users’ sketching habits and symbol knowledge may result in diverse patterns. That is, a unambiguous mapping language is necessary to interpret drawings into machine-readable instructions. The language we designed exhibits three fundamental features, each with its own references.

\begin{itemize}[leftmargin=0.5cm]
\item The first is the use of colors to distinguish between the activities inspired by a laser cutter. When performing tasks that involve varying levels of power or speed, laser cutting employs a technique known as color-mapping \cite{colormapping}, which allows users to assign different settings to different regions of an image depending on the color.
\item The second step is the operational procedure. As noted above, the CAD processes consist of three stages: blueprints, toolpaths, and previews. We believe that users can utilize \systemName similarly. That is, the user first draws a blueprint with a specific color, and then specifies the functions by drawing a syntax with other colors.
\item The third is the types of products that can be made by a three-axis CNC machine. The three-axis CNC machines, with the mill bit moving from top to bottom, are used for vertical subtraction. Consequently, they can be used to carve contour patterns and pockets or holes of various shapes, but only for limited 3D engraving (cut a part with flat features of varied depth) \cite{2.5D}. Based on these three types, we choose the three colors to cover all conceivable cutting results.
\end{itemize}

To summarize, we chose the three existing colors (purple, red, and green) because of their significant variances in HSV values, which improves the color identification accuracy. Pen colors were used to symbolize the various functions (Figure~\ref{fig:2}c). For example, a purple pen indicates a blueprint. The red and green pen symbols depict the tool path instructions. Symbols can be systematized to ensure rigidity, and people typically use them in their daily lives. Past research has explored the strong expressiveness of freehand annotation~\cite{song2006modelcraft}, and certain symbols (such as crosses) have well-known meanings that allow users to draw instructions naturally. Inspired by the color mapping of laser cutting, the color language of \systemName can be customized by the user.

\paragraph{What are the applications (Figure~\ref{fig:2}, part d)?}

CAD is the dominant tool for planning complex 3D architectures with various elements, designing millimeter-accurate gears, and mass-producing goods for standard models. We do not intend to replace general-purpose CAD with \systemName. The proposed system should have a specific scope for use in the targeted applications. Two promising applications are identified. The first is the creation of customized artifacts. This can be a one-of-a-kind artifact \cite{zoran_freed_2013} recording improvisational thoughts \cite{tian_matchsticks_2018} or handwritten identifications. \systemName also reduces the skill level required for its use. Children can even create models quickly by drawing on wood. The second is the fabrication requirements for measuring, scanning, or constructing real items as references during the manufacturing process. Examples include joint customization and furniture modification.

\begin{figure*}[!h]
    \centering
    \begin{subfigure}{0.25\textwidth}
        \centering
        \includegraphics[width=0.95\linewidth]{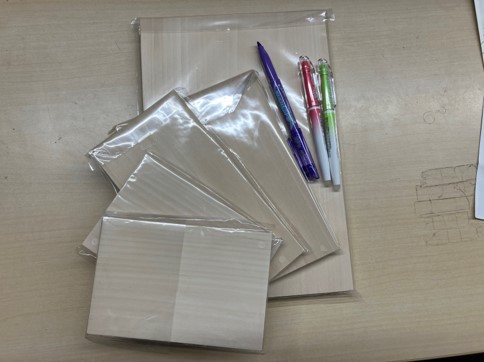}
        \caption{Tools}
    \end{subfigure}%
    \begin{subfigure}{0.25\textwidth}
        \centering
        \includegraphics[width=0.95\linewidth]{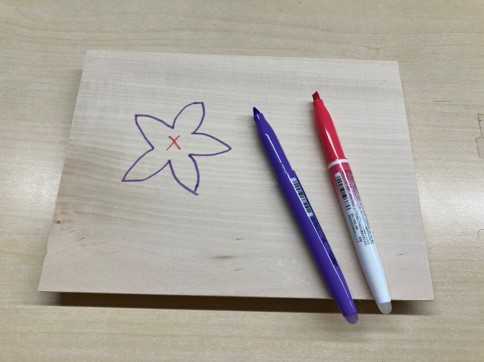}
        \caption{Drawing}
    \end{subfigure}%
    \begin{subfigure}{0.25\textwidth}
        \centering
        \includegraphics[width=0.95\linewidth]{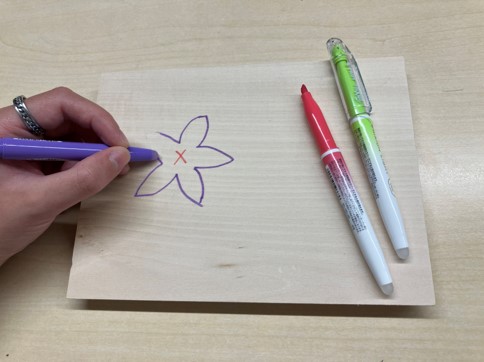}
        \caption{Revising}
    \end{subfigure}%
     \begin{subfigure}{0.25\textwidth}
        \centering
        \includegraphics[width=0.95\linewidth]{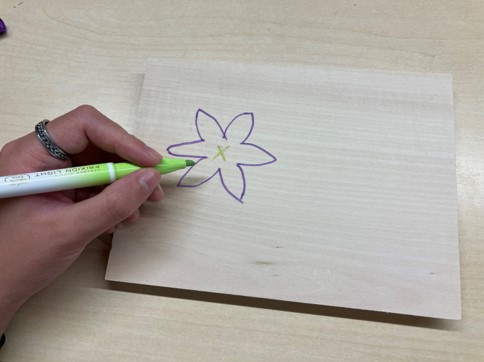}
        \caption{Final sketch}
    \end{subfigure}%
    \caption{Various tools, including (a) several sizes of wood and three erasable color pens. If the user is dissatisfied with their drawing (b), they may erase it (c) and redraw it (d).}
    \Description{Various tools, including (a) several sizes of wood and three erasable color pens. If the user is dissatisfied with their drawing (b), they may erase it (c) and redraw it (d).}
    \label{fig:3}
    \end{figure*}

\begin{figure*}[!h]
\includegraphics[width=0.9\textwidth]{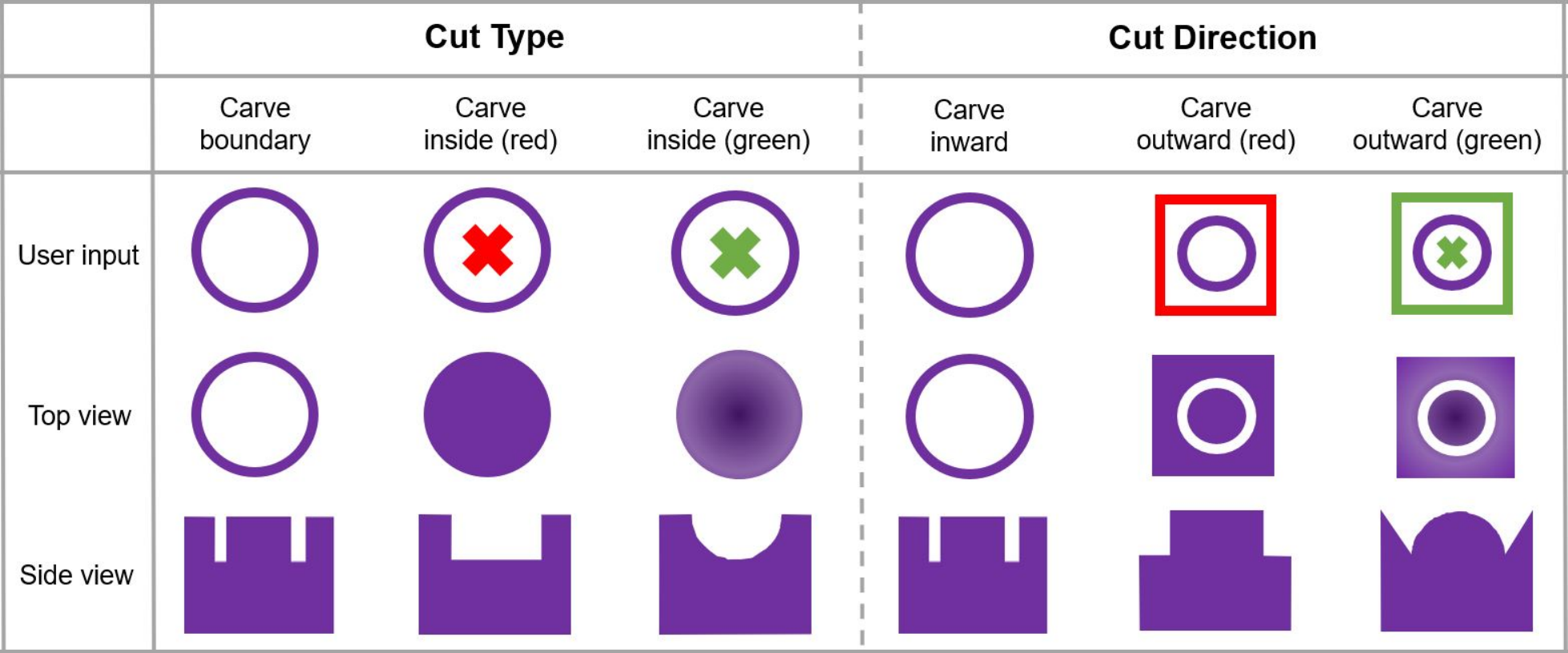}
\caption{Concept of drawing language illustrating how various colors and symbols can be utilized to accurately communicate cut intent and the method. Left panel: Top view of symbols describing the cut type and cut direction. Right panel: Side view of the difference in output between using the red pen and the green pen.}
\Description{Concept of drawing language illustrating how various colors and symbols can be utilized to accurately communicate cut intent and the method. Left panel: Top view of symbols describing the cut type and cut direction. Right panel: Side view of the difference in output between using the red pen and the green pen.}
\label{fig:4}
\end{figure*}

\subsection{Technical Flow and Requirements}

We performed a fast prototyping exploration to identify user tasks and workflows. We constructed a basic low-fidelity prototype. We used \textit{VCarve Pro}’s tracing tool to instruct the \textit{ShopBot} CNC machine\footnote{\url{https://shopbottools.com/products/desktop/}} to cut along the user’s hand-drawn curve. The procedure consists of 11 steps, divided into three parts.

\begin{enumerate}[leftmargin=0.5cm]
   \item[(a)] Preparation to setup the digital representation:
   \begin{itemize} 
     \item \textit{Draw the "Cross" pattern on the wood and save it as a bitmap}
     \item \textit{Crop the image to remove the background}
     \item \textit{The wood is measured and its height, length, and width are used to set up a job file}
   \end{itemize}
   \item[(b)] Digital manipulation on CAD:
   \begin{itemize} 
     \item \textit{Import the cropped bitmap to the canvas}
     \item \textit{The built-in color-tracing feature is used to isolate the vector with the registered black color}
     \item \textit{Adjust the corner fit and noise filter for the isolated vector}
     \item \textit{Select the V-Carve/Engraving toolpath and specify the depth for generating a G-code file}
   \end{itemize}
   \item[(c)] Physical space registration:
   \begin{itemize} 
     \item \textit{Fix the wood to the machine bed}
     \item \textit{Calibrate the three axes to make the lower-left corner of the wood a zero point}
     \item \textit{The V-bit milling tool was selected, and the CNC machine was activated by loading the G-code file}
     \item \textit{After the wood is cut, we remove it from the CNC machine}
   \end{itemize}
\end{enumerate}

Some tasks, such as planning blueprints and toolpaths, are related to the user’s design intent only in part (b). Some tasks are trivial but mandatory for every cut. Part (a) involve recording digital data to create a digital representation. In steps part (c), the virtual model is calibrated before the computed results are returned to a physical space of the same size. When using this fast prototyping method or the standard CAD/CAM approach, users should register manually. We intend to create an automatic registration module to internalize the setup and calibration.

If the cut preview by CAD differs from what we intend, then we repeat steps in part (a) to achieve the desired output. Therefore, we believe that even if all interactions occur in a physical space (i.e., on wood), \systemName still requires an additional screen to provide a cut preview. We identified three modules to complete this implementation.

\begin{itemize}[leftmargin=0.5cm]
\item The \textbf{\textit{registration module}}, which aligns the physical working space with the virtual model.
\item The \textbf{\textit{language module}} enables the CNC machine can interpret user’s drawn patterns precisely.
\item The \textbf{\textit{interface module}}, which displays the cut preview to the user for validation.
\end{itemize}

\begin{figure*}[!h]
\includegraphics[width=1\textwidth]{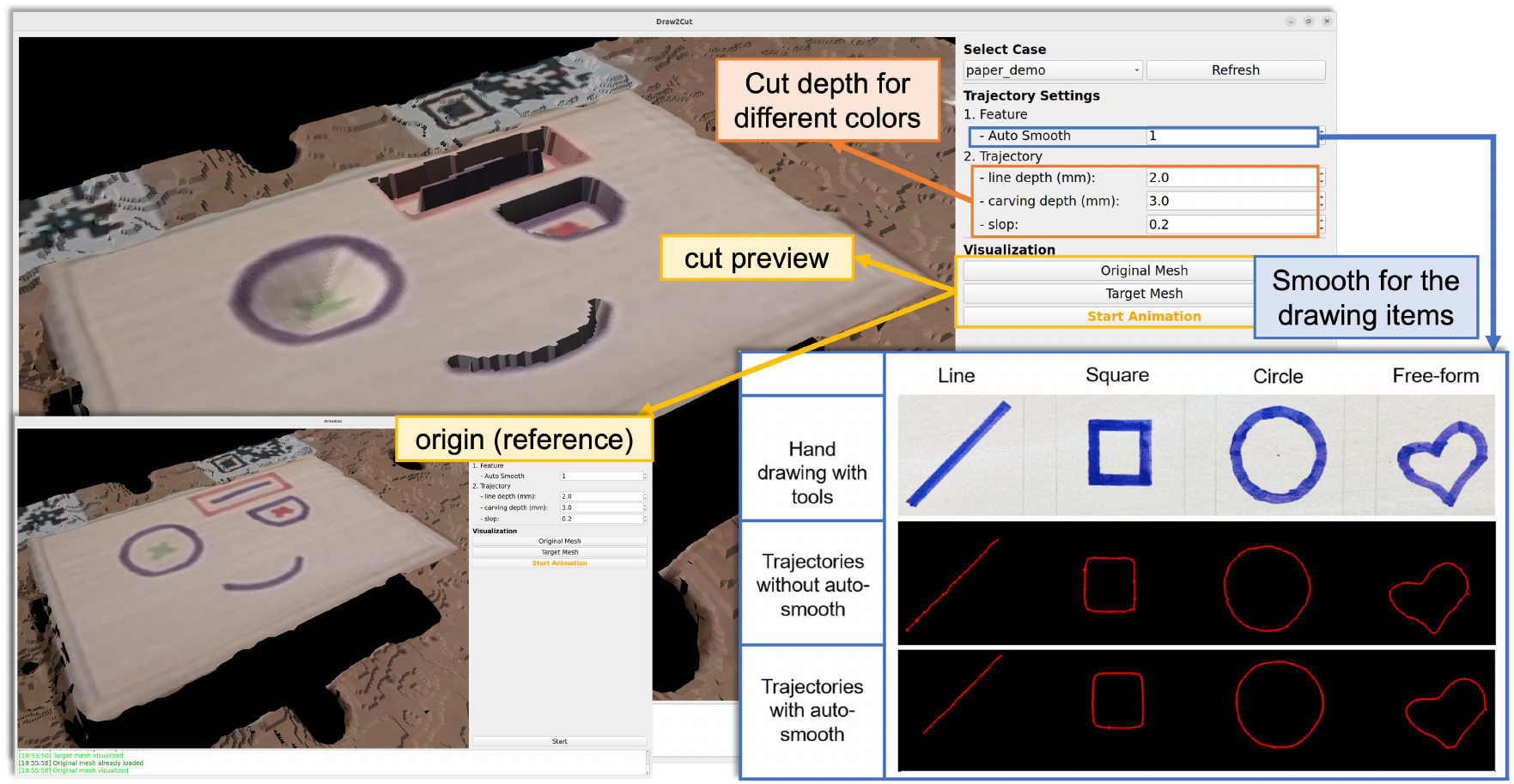}
\caption{Visualization interface for users to specify the cut depth (the orange box on the right) and smooth size (the blue box) for each item, providing users with three types of visualizations: reconstructed mesh based on the collected point cloud data (interface on the left-bottom), the target mesh that shows the result, and an animation that explains the cut procedure.}
\Description{Visualization interface for users to specify the cut depth (the orange box on the right) and smooth size (the blue box) for each item, providing users with three types of visualizations: reconstructed mesh based on the collected point cloud data (interface on the left-bottom), the target mesh that shows the predicted result, and an animation that explains the cut procedure.}
\label{fig:5}
\end{figure*}

\section{\systemName Workflow}

The key feature of \systemName is that users can sketch on wood directly. Consequently, instead of a typical user interface on a computer screen or drawing pad, the target material itself serves as the user interface for our system (Figure~\ref{fig:1}, left). \systemName has a primary input method (Figure~\ref{fig:3}a) involving three erasable pens, and a secondary method involving a desktop interface (Figure~\ref{fig:5}) for validation. The three erasable pens (purple, red, and green) allow users to draw the design intent and cut out the instructions. The visualization interface displayed on the desktop monitor shows cut preview, trajectory, and customization options. The fabrication of the product involved four steps.

\paragraph{\textbf{Step 1: User draws the contour of the design intent on wood with a drawing pen (purple).}}
The user just needs to sketch the desired shape of carving in this step (Figure~\ref{fig:3}b). The user can start by free-drawing or making auxiliary lines on the wood. The pen was inked (clear enough to be detected) and erasable. Users can erase measured lines or incorrectly draw lines (Figure~\ref{fig:3}c and ~\ref{fig:3}d).

\paragraph{\textbf{Step 2: User tells the machine how to cut using the behavior pens (red and green).}} 

The behavior pens produce two types of instructions.

\begin{itemize}[leftmargin=0.5cm]
\item The first is the cut type (Figure~\ref{fig:4}, left). If it is a letter or a curve, then it should be engraved (i.e., the boundary is carved). If the user draws a closed form, such as a circle or polygon, then the command should specify whether only the outline should be engraved or all portions of the shape should be cut (pocket toolpath). To identify the pocket (i.e., carve inside), we used the behavior pen to draw a "cross" inside the closed form.
\item The second is the cutting direction (Figure~\ref{fig:4}, right). Occasionally, we intend to carve an inward pattern that entails creating sunken elements in a material. Simultaneously, to create a seal stamp, the name should be carved outward, which requires the material to be removed to form convex separate shapes. If the user intends to carve an object outward, then they can use a behavior pen to draw a closed shape around the target. When the system recognizes this, it removes all the wood within the closed shape, except for purple strokes (similar to a \textit{NOT} gate).
\end{itemize}

\begin{figure*}[!h]
    \centering
    \begin{subfigure}{0.5\textwidth}
        \centering
        \includegraphics[width=0.98\linewidth]{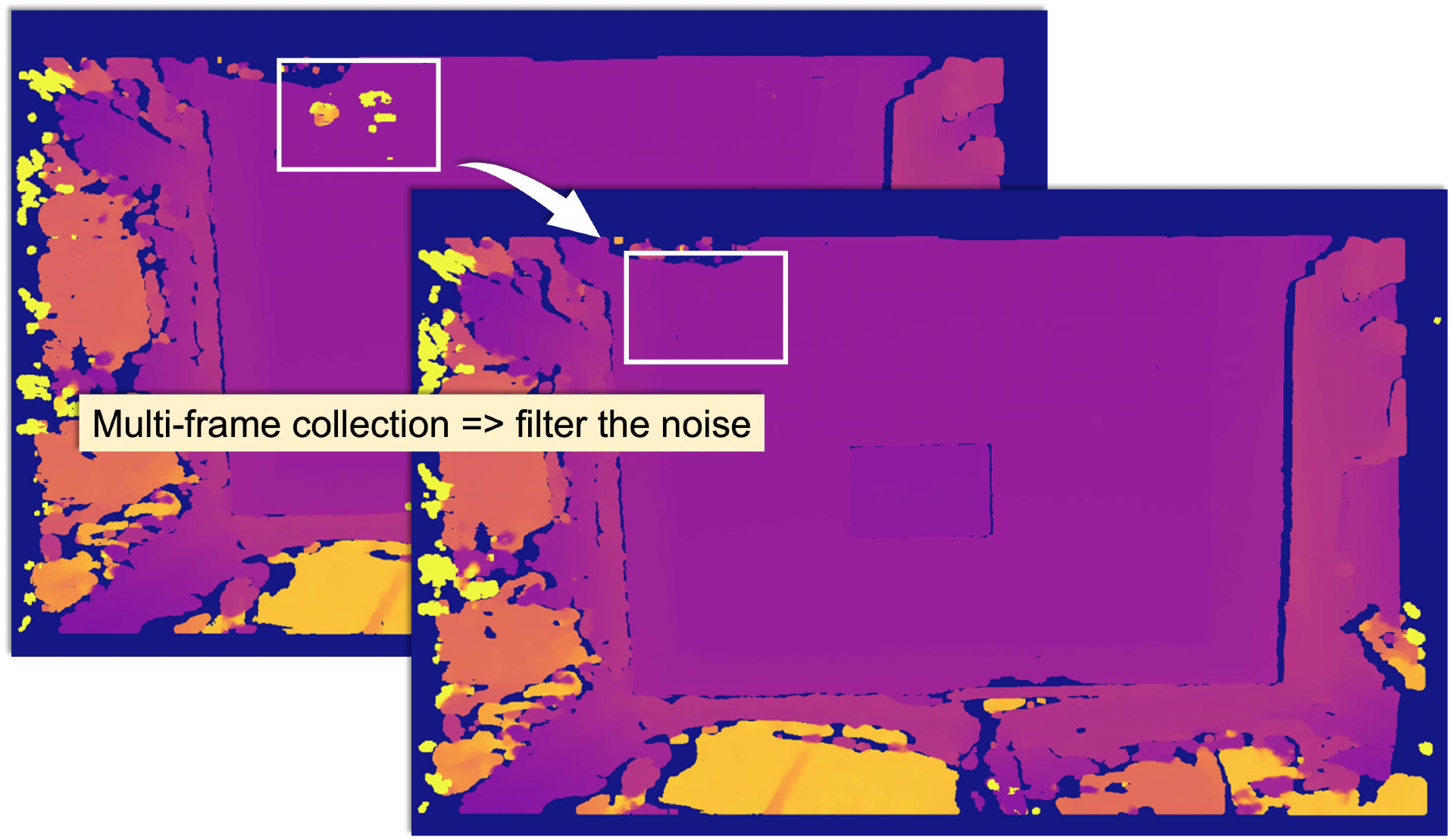}
        \caption{Multi-frame data collection to filter noise in white box}
    \end{subfigure}%
    \begin{subfigure}{0.5\textwidth}
        \centering
        \includegraphics[width=0.98\linewidth]{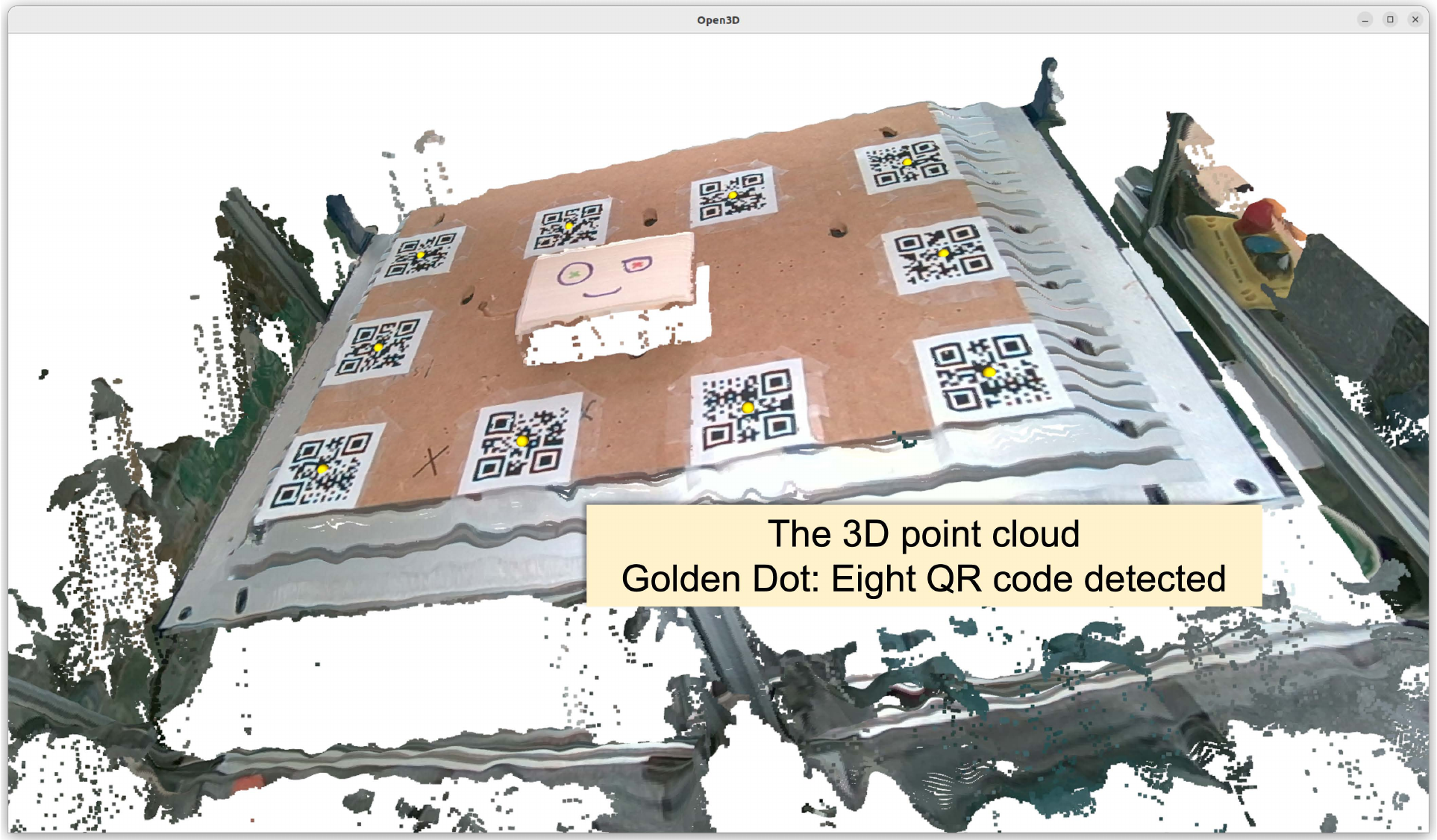}
        \caption{3D point cloud and virtual coordinate construction}
    \end{subfigure}%
    \newline
    \begin{subfigure}{0.445\textwidth}
        \centering
        \includegraphics[width=0.98\linewidth]{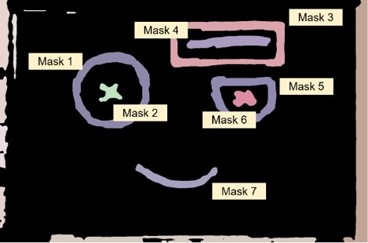}
        \caption{Mask extraction from the detected surface}
    \end{subfigure}%
    \begin{subfigure}{0.555\textwidth}
        \centering
        \includegraphics[width=0.98\linewidth]{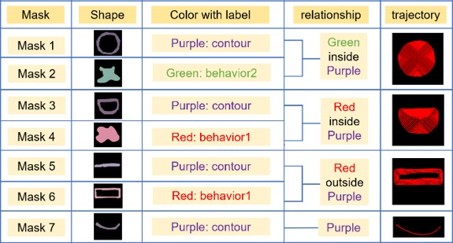}
        \caption{Mask analysis to determine where to cut}
    \end{subfigure}%
    \caption{Technical processes for data collection, virtual space construction, image interpretation, and trajectory planning.}
    \Description{Underlying technical process from data collect (a), virtual space construction (b), image interpretation (c), and trajectory planning (d).}
    \label{fig:6}
    \end{figure*}

We use red and green behavior pens. Both pens share the same symbols but create different results. In the graving process, the red pen indicates a hole with vertical walls, and the green pen represents a hole with curved walls. For example, if a red cross is drawn inside a purple circle, then the CNC cuts a cylindrical hole (Figure~\ref{fig:4}, right). If we draw a green cross inside the purple circle, then the CNC is cut into a concave hemisphere. Similarly, drawing a red square outside the purple circle causes the CNC to cut a circle that is convex relative to its surrounding. If we draw a green square outside the purple circle, then the CNC creates a mound. The curvature of the cut surface was positively correlated with the cutting depth, as specified by the user in the next step.

The user then placed the material on a machine bed in the workspace. The depth camera mounted on the top glass collects data, which is then used by \systemName to analyze the user’s lines and symbols to interpret the fabrication intent and construct a cutting path.

\paragraph{\textbf{Step 3: Visualization interface.}}
The user sets the parameters and makes the preview validation. The interface contains a setting panel for the two parameters and a visualization canvas.

\begin{enumerate}[leftmargin=0.5cm]
  \item \textbf{Parameter 1: Auto-smooth}. This function can remove noise (e.g., hand shaking and ink seeping) from handwriting and introduce smooth lines for features, such as signatures. Furthermore, if users wish to create a standard pattern, such as a straight line or a regular pattern, they first sketch using the support tools (ruler, protractor, compass) and then select a target component and click "auto-smooth." (Figure~\ref{fig:5}, blue). However, this method cannot be used to draw the rectified shapes (perpendicular or straight) accurately. In the future, we plan to add other supported functions (e.g., regularization).
  \item \textbf{Parameter 2: Cut depth}. Unlike abstract elements, such as irregular curves or shapes, which are difficult to convey verbally, depth data are concrete and can be measured in millimeters. In addition, depth is closely related to previews. We allowed the users to adjust the depth for different cut types while viewing the preview at various depths. (See Figure~\ref{fig:5}, light orange).
  \item \textbf{Visualization}. The interface provides three types of visualization. The user can preview the visualization by clicking on each button according to their preferences (see Figure~\ref{fig:5}, yellow part): a) \textbf{Origin}: source image as a reference. b) \textbf{Target}: cut preview displays the expected results in 3D view. c) \textbf{Animation:} rendered video conveys the user cut procedure. These visualizations help determine whether \systemName interprets the user’s design intent correctly, pitches the cut position precisely, and plans the cut path reasonably. When users are satisfied with the preview visualization, they proceed to confirm.
\end{enumerate}

\paragraph{\textbf{Step 4: Fabrication.}}
After confirming the preview, the system updates the generated G-code, uploads it to the CNC system, and automatically moves to cut the output. After cutting is complete, the user can assemble, paint, or sand-cut pieces to complete the final product.

\begin{figure*}[!h]
\includegraphics[width=0.9\textwidth]{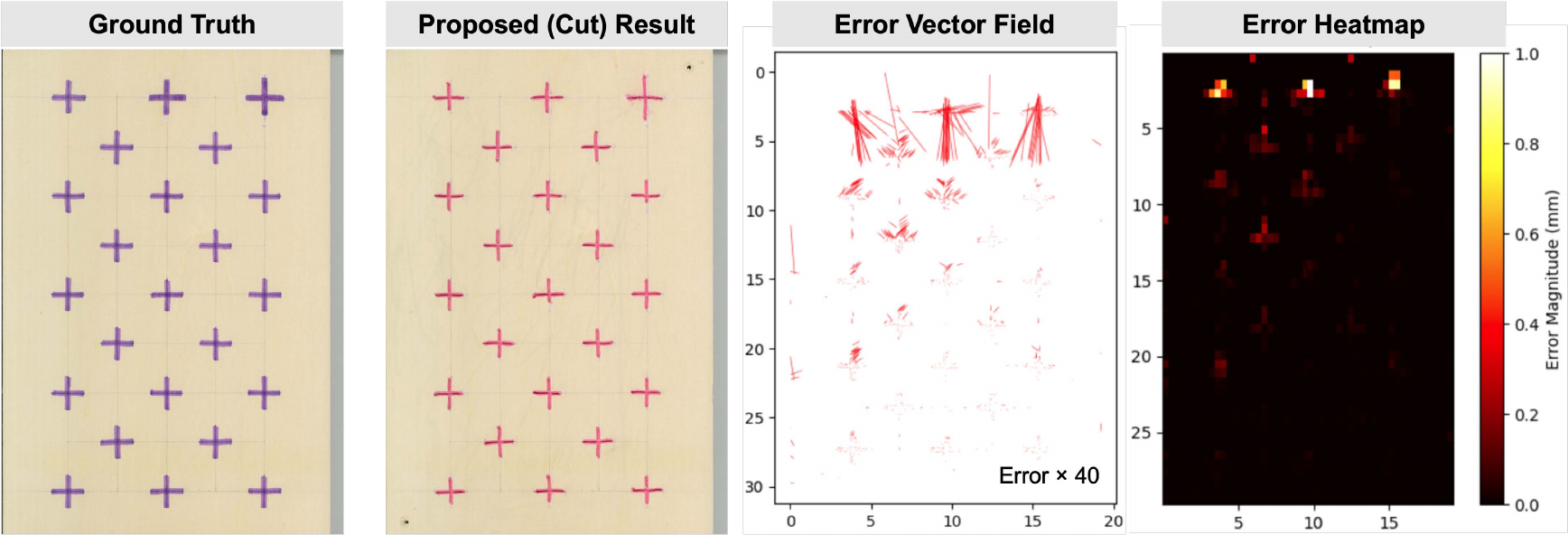}
\caption{The first two scanned figures from the hand-drawn and machine-cut result. The following two images are the error distribution shown as a vector field and heatmap. In the vector graph, we enlarged the length of the vector by 40 times to make it easier for readers to see. If people look at the heatmap, the maximum error will not exceed 1 mm (the white pixel).}
\Description{The first two scanned figures from the hand-drawn and machine-cut result. The following two images are the error distribution shown as a vector field and heatmap. In the vector graph, we enlarged the length of the vector by 40 times to make it easier for readers to see. If people look at the heatmap, the maximum error will not exceed 1 mm (the white pixel).}
\label{fig:7}
\end{figure*}

\section{Implementation}

\subsection{Setup}

First, we set up a physical workspace to capture the patterns drawn on wood. We identified the workspace using ten QR code patches. Users can draw their design intent on any physical materials (e.g., a comfortable sofa or office desk). They then placed the drawn wood within the range of ten QR codes. An activated depth camera fixed above the workspace captured frames in two modalities (depth and color). To ensure a higher accuracy, we captured several depth frames, calculated the average, and filtered out the noise, as depicted in the white box in Figure~\ref{fig:6}a.

\subsection{Registration}

The registration process aligns the virtual space with the real-world coordinates. The virtual space features a standard orthonormal basis in 3D space. The collected depth map and color images were transformed into a colored point cloud, which maintained the same scale as the real-world coordinates (Figure~\ref{fig:6}b). The centers of each QR code were identified and used to define the XY plane and axis ranges in the virtual space. We computed the normal vector of the plane to establish the Z-axis. Similarly, we identified the corresponding center points in the point cloud, which in turn defined the real-world coordinates. Finally, we calculated the transformation matrix between the real and virtual spaces. 

\subsection{Surface Extraction}

The next stage involved retrieving the positions of the physical material pieces and extracting the texture of the surface. To eliminate irrelevant and noisy data points, we maintained the points within the XY-axis range and limited the Z-axis range within 5 to 160 mm. Starting from the top of the z-axis range, we searched one step at a time, with each step being 0.5 mm. When we reached a threshold with more than 50\% of the total points above it, we defined it as the surface z-value. The collected points are used to form a 2D surface using a warping algorithm. We upscaled it by a factor of 10 for subsequent processing.

\subsection{Drawn Pattern Interpretation} 

Using an auto-threshold algorithm, we extracted masks for the colored sections and calculated the representative colors within each masked region (Figure~\ref{fig:6}c). We prepared a dictionary that associated each color with a specific intention (Figure~\ref{fig:6}d). Next, we determined the centerline of each mask and categorized it into loops and lines. This information enabled us to determine topological relationships.

The cutting depth for each point in the region is a function of the distance from the point to the region's boundary. Different carving patterns are created by applying various kernel functions and setting a depth limit to map the distance to the carving depth. We used a steep slope to map the distance-to-depth linearly for red pen cut. We applied a user-determined slope for green pen cut. Given the carving depth map, the cut was processed from coarse to fine. Every cutting step subsequently identified the uncut mask within the depth map. We planned an inward-cutting trajectory using contour extraction and an erosion algorithm.

\subsection{Cut Preview Rendering}

Based on the collected point cloud data, we reconstructed the mesh, shown as "Origin” (Original mesh) in Figure~\ref{fig:5}. Combining this with the carving depth obtained, as discussed in the previous section, we estimated the target mesh after the cutting process, depicted as "Target" in Figure~\ref{fig:5}. Then, using the planned trajectory and assuming that the circular region around each point on the trajectory is trimmed, we interactively updated the mesh. This produced the animation preview of the cutting trajectory, as shown in the "Animation" in Figure~\ref{fig:5}. These crucial previews help users validate whether the algorithm generates a reasonable G-code, ensuring that the CNC machine will be cut according to user preferences.

\begin{figure*}[!htb]
\includegraphics[width=1\textwidth]{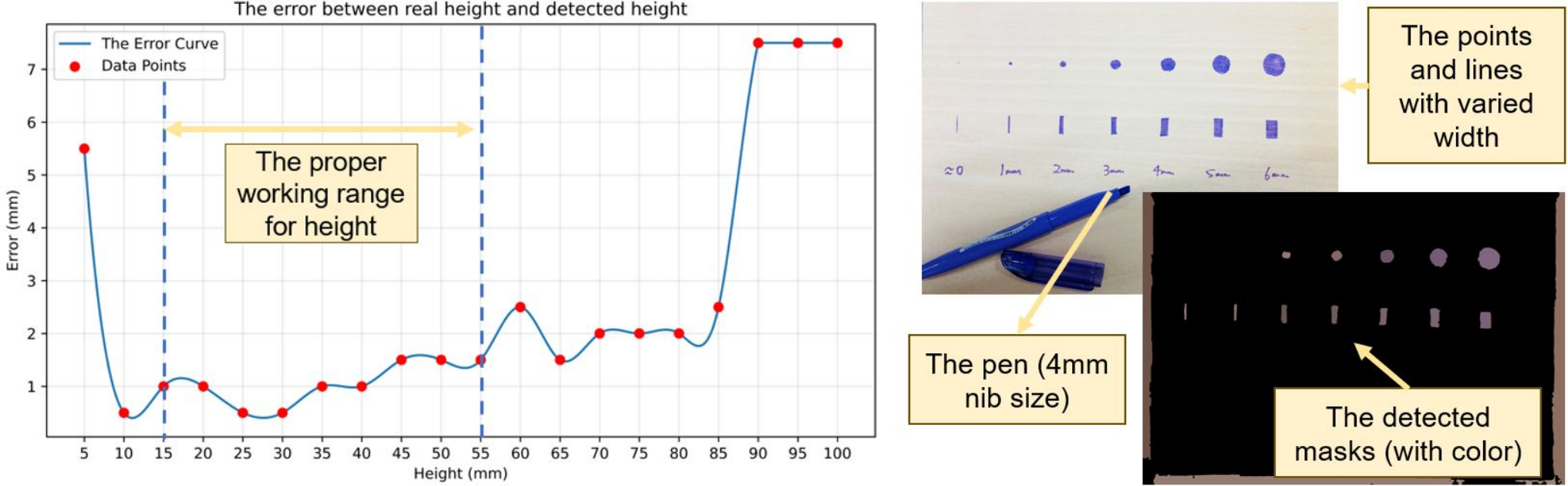}
\caption{Left: a line graph describes how the height of the wood surface influences the error of detection. Right: the determination of the minimal feature size, both in line and point.}
\Description{Left: a line graph describes how the height of the wood surface influences the error of detection. It shows that the best working height is with 15-55mm. Right: the determination of the minimal feature size, both in line and point.}
\label{fig:8}
\end{figure*}

\section{Technical Evaluation}

\subsection{Virtual–Physical Alignment Accuracy}

The physical material should fit within the area of the quadrilateral formed by the center points of the four QR codes at the corners. The experimenters measured the actual area of this space; the x-axis was 300 cm and the y-axis was 527 cm. The x- and y-axes of the constructed virtual space were 296 cm and 524 cm, respectively. Therefore, the error along the x-axis was compressed inward by 1.3\%, and the error along the y-axis was compressed inward by 0.57\%. We conducted a technical evaluation to obtain the detailed error distribution results. This task involves using a CNC machine to cut a mesh structure (e.g., grid).

First, the material was prepared (Figure~\ref{fig:7}, two images to the left). A $5 \times 9$ grid was drawn on the wood, and crosses were drawn from top to bottom at 23 points. The image was scanned and saved as the ground truth. We extracted 23 crosses using \systemName software. We color the cut areas pink, scan them again using the same machine, and save them as the proposed (cut) results. We compared and analyzed the two images to obtain a vector error plot and error heatmap (Figure~\ref{fig:7}, right). The error distributions and positions relative to the depth camera were compared. The results revealed that the smaller the angle between the line connecting the point on the machine tool and depth camera compared with the horizontal plane, the greater the error. Thus, the error was the smallest at the point directly below the depth camera (the angle was approximately 90°). The error is not larger than 0.5 mm around the center, and the cut accuracy inside the workspace is sufficient for fabrication.

\subsection{Detection Limitations}

The fundamental limitation of the height estimation is the working range of the depth camera (\textit{Intel RealSense D435}). The depth camera was positioned 60 cm away from the machine bed. Within a range of (0-10 cm), we increased the height of the wood piece by five millimeters and recorded the measured and detected heights. We subtracted the two heights to calculate the inaccuracy for this wood-block thickness and plotted it (Figure~\ref{fig:8}, left) to show how the material thickness affects the accuracy. In the demonstration described in the next section, we placed the wood surface at a height of 15–55 mm and manually adjusted the z-axis offset to improve the accuracy. Owing to the differences in depth camera types and workspace settings, an appropriate working height should be determined for each working environment. The plot obtained in this experiment is exclusively applicable to the environment, as shown in Figure~\ref{fig:1}.

The number of items drawn on the material should be greater than the minimum feature size. We evaluated points and lines with different radii and line widths (Figure~\ref{fig:8}, right). The results indicate that using a pen with a stroke width greater than 4 mm improves the color detection accuracy. If the line is too thin, it cannot be recognized (owing to the depth camera resolution), and is easily influenced by the backdrop color. In the following experiment, we chose a pen with a 4 mm nib.

\section{System Demonstration}

\subsection{Basic Advantages}

This section highlights the two major applications that show the advantages compared to CAD that \systemName offers.

\paragraph{Aspect 1: Fabrication with reference objects.} 

\begin{figure*}[!htb]
\includegraphics[width=1\textwidth]{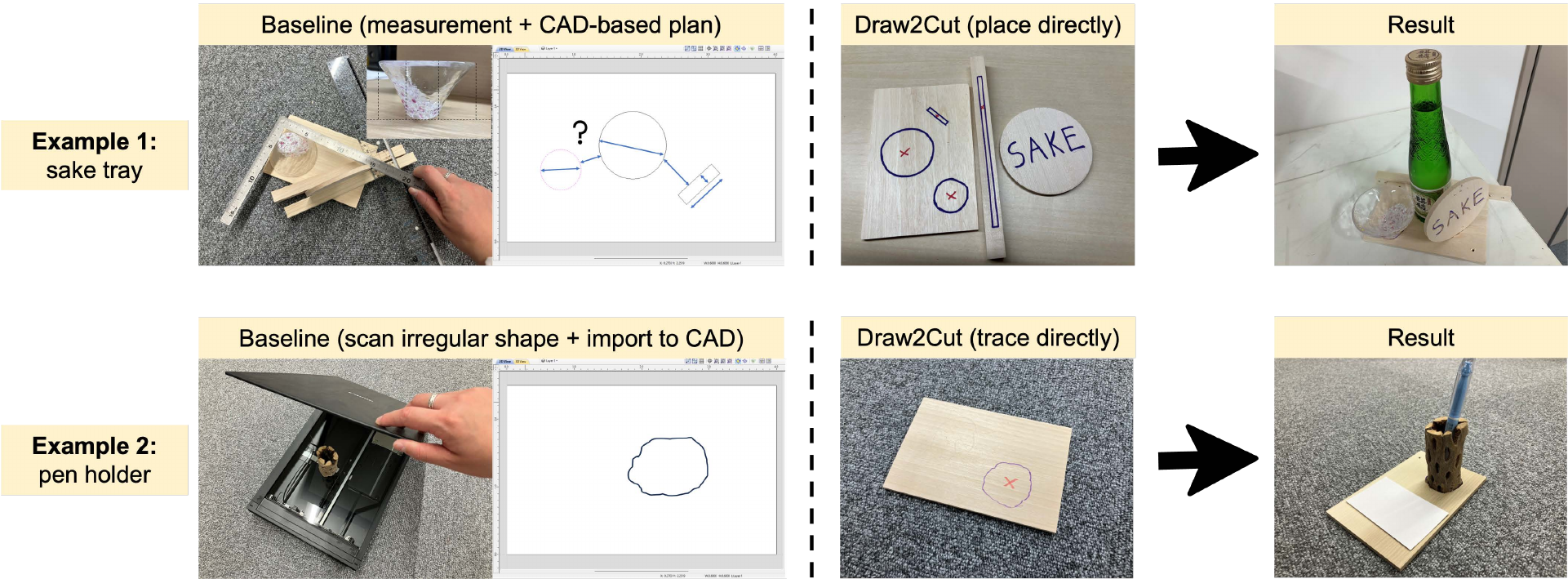}
\caption{Advantage of aspect 1: Baseline (CAD-based design) vs \systemName in the preparation step. \textbf{in example 1,} because the mouth and bottom of a sake glass are different in size, in order to fit both of them in a small space, CAD users need to measure different data. However, \systemName users only need to group all of the objects, put them in their proper places, and draw on them directly.}
\Description{Advantage of aspect 1: Baseline (CAD-based design) vs \systemName in the preparation step. \textbf{in example 1,} because the mouth and bottom of a sake glass are different in size, in order to fit both of them in a small space, CAD users need to measure different data. However, \systemName users only need to group all of the objects, put them in their proper places, and draw on them directly.}
\label{fig:9}
\end{figure*}

The first example illustrates the fabrication required for measurement using actual products as references, such as assembly tasks. As shown in Figure~\ref{fig:9}, we created a Japanese sake tray. There are multiple holes in the tray to hold various items. It is difficult to arrange the sizes and positions of cuts using CAD. For example, sake glass has an irregular shape. The direct measurement of wood can ensure that the distance between each hole is reasonable. The second example illustrates an irregularly shaped carving. A normal CAD interface requires users to scan an irregular object first and subsequently input the scanned result into a computer for conversion into a vector. Drawing irregular shapes as vectors in software is difficult, but sketching them manually is easy. The \systemName system allows users to sketch the irregular shape of the wood.

\begin{figure*}[!htb]
    \centering
    \begin{subfigure}{0.23\textwidth}
        \centering
        \includegraphics[width=0.95\linewidth]{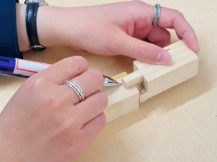}
        \caption{Auxiliary line is drawn for a joint with two legs}
    \end{subfigure}%
    \hspace{0.2cm} %
    \begin{subfigure}{0.23\textwidth}
        \centering
        \includegraphics[width=0.95\linewidth]{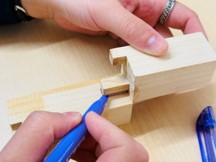}
        \caption{Rectangle required to cut and insert the long leg is traced}
    \end{subfigure}%
    \hspace{0.2cm} %
    \begin{subfigure}{0.23\textwidth}
        \centering
        \includegraphics[width=0.95\linewidth]{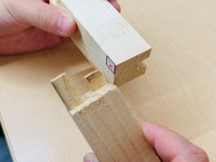}
        \caption{Repeat previous step to insert the short leg after flipping}
    \end{subfigure}%
    \hspace{0.2cm} %
    \begin{subfigure}{0.23\textwidth}
        \centering
        \includegraphics[width=0.95\linewidth]{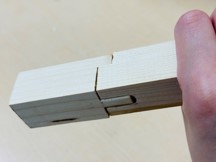}
        \caption{Cut is finished and the edge joint is completed}
    \end{subfigure}%
    \hspace{0.2cm} %

    \begin{subfigure}{0.23\textwidth}
        \centering
        \includegraphics[width=0.95\linewidth]{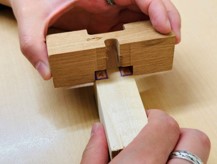}
        \caption{Two squares are traced to plug the lower part}
    \end{subfigure}%
    \hspace{0.2cm} %
    \begin{subfigure}{0.23\textwidth}
        \centering
        \includegraphics[width=0.95\linewidth]{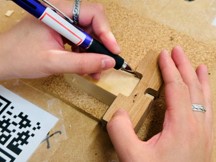}
        \caption{Half of the square is cut and inserted into the joint}
    \end{subfigure}%
    \hspace{0.2cm} %
    \begin{subfigure}{0.23\textwidth}
        \centering
        \includegraphics[width=0.95\linewidth]{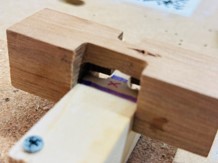}
        \caption{Rectangular required to cut for the upper part}
    \end{subfigure}%
    \hspace{0.2cm} %
    \begin{subfigure}{0.23\textwidth}
        \centering
        \includegraphics[width=0.95\linewidth]{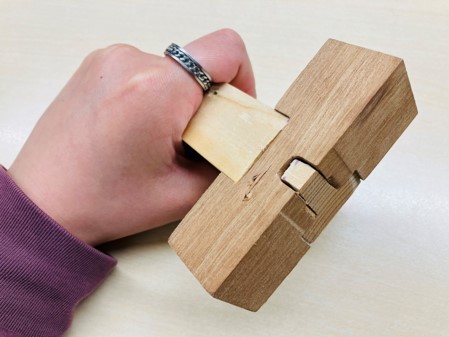}
        \caption{Cut is finished and the T-joint is completed}
    \end{subfigure}%
    \hspace{0.2cm} %
    \caption{Edge joint (first row), and T-joint (second row) made by \systemName.}
    \Description{The edge joint (first row) and T-joint (second row) made by \systemName. In the edge joint, the user first draws an auxiliary line for a joint with two legs, then traces the rectangle needed to cut to insert the long leg. After that, the user flips and does the same for inserting the short leg and finishes the cut and completes the edge joint finally. In the T-joint, the user first traces two squares to plug the lower part, then cuts them first to insert half into the joint. Later, the user traces the rectangular need to cut for the upper part and finishes the cut and completes the T-joint.}
    \label{fig:10}
    \end{figure*}

\paragraph{Aspect 2: Creating customized artifacts. }

We propose that people can create custom stamps using their signatures, preferred characters, or company logos. Second, there are some classic wooden games, such as the wooden maze, jigsaw puzzle, and \textit{Mancala}. Users can utilize these classic game concepts as the basis for designing their tracks or patterns. First, we performed an internal group test for the \systemName system. In our group demonstration (Figure~\ref{fig:1}, right), we designed the Sakura group logo for our system, and plotted a wooden ball maze using a customized route.

\begin{figure*}[!htb]
    \centering
    \begin{subfigure}{0.25\textwidth}
        \centering
        \includegraphics[width=0.95\linewidth]{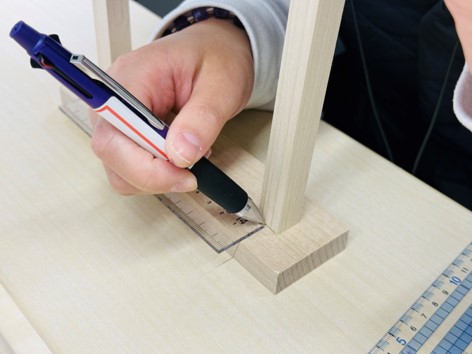}
        \caption{Measure}
    \end{subfigure}%
    \begin{subfigure}{0.25\textwidth}
        \centering
        \includegraphics[width=0.95\linewidth]{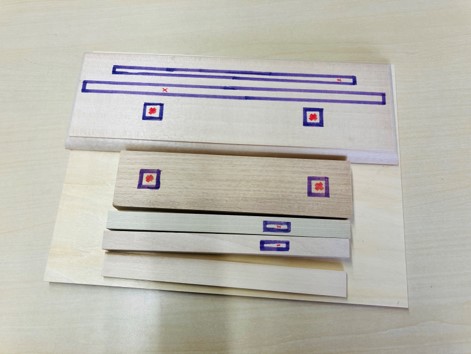}
        \caption{Trace}
    \end{subfigure}%
    \begin{subfigure}{0.25\textwidth}
        \centering
        \includegraphics[width=0.95\linewidth]{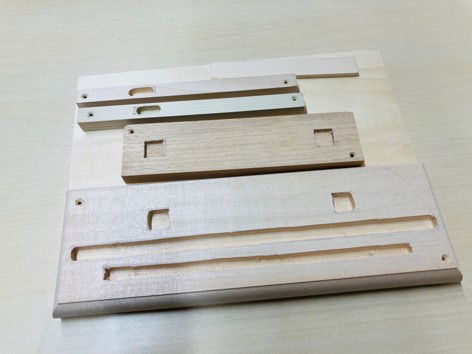}
        \caption{Place and cut}
    \end{subfigure}%
    \begin{subfigure}{0.25\textwidth}
        \centering
        \includegraphics[width=0.95\linewidth]{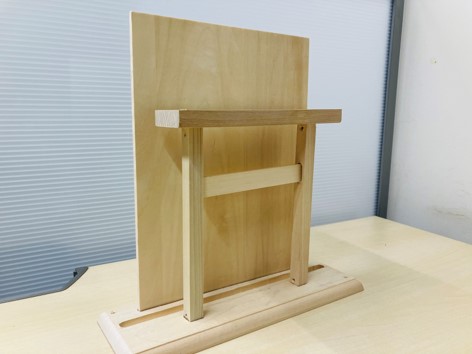}
        \caption{Assemble}
    \end{subfigure}%
    \caption{Torri was created using \systemName. First, the cut location is identified and an auxiliary line is drawn with a ruler; the cut area in purple is then traced and filled with red crosses. The user places the wood pieces on the machine, and the CNC automatically cuts and completes the assembly.}
    \Description{Torri was created using \systemName. First, the cut location is identified and an auxiliary line is drawn with a ruler; the cut area in purple is then traced and filled with red crosses. The user places the wood pieces on the machine, and the CNC automatically cuts and completes the assembly.}
    \label{fig:11}
    \end{figure*}

\begin{figure*}[!htb]
    \centering
    \begin{subfigure}{0.20\textwidth}
        \centering
        \includegraphics[width=0.95\linewidth]{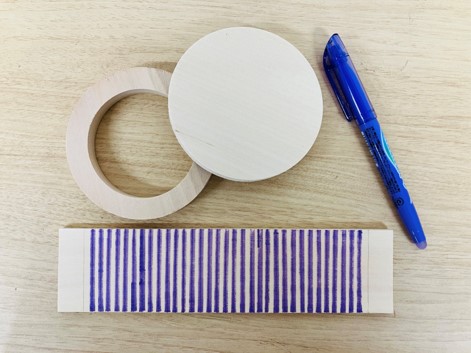}
        \caption{Draw lines}
    \end{subfigure}%
    \begin{subfigure}{0.20\textwidth}
        \centering
        \includegraphics[width=0.95\linewidth]{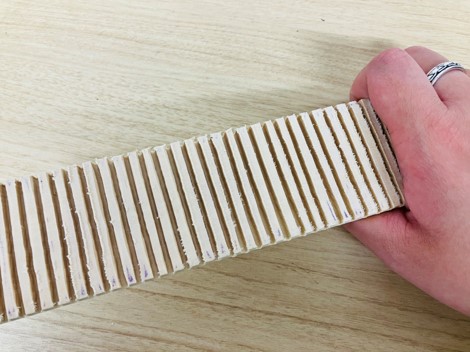}
        \caption{Cut by \systemName}
    \end{subfigure}%
    \begin{subfigure}{0.20\textwidth}
        \centering
        \includegraphics[width=0.95\linewidth]{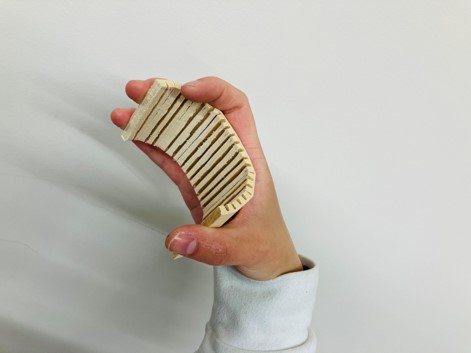}
        \caption{Bent case 1}
    \end{subfigure}%
    \begin{subfigure}{0.20\textwidth}
        \centering
        \includegraphics[width=0.95\linewidth]{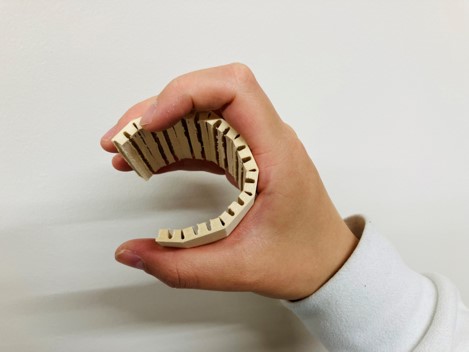}
        \caption{Bent case 2}
    \end{subfigure}%
     \begin{subfigure}{0.20\textwidth}
        \centering
        \includegraphics[width=0.95\linewidth]{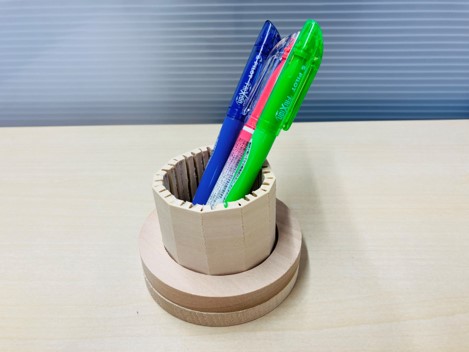}
        \caption{Small item container}
    \end{subfigure}%
    \caption{Kerf made by \systemName. User first draws lines using a ruler with 5 mm distance, and places it to the machine. The \systemName cut result can be bent by different degrees to achieve varied shapes, being the fence for pen holder or a stand for small items.}
    \Description{Kerf made by \systemName. User first draws lines using a ruler with 5 mm distance, and places it to the machine bed. The \systemName cut result can be bent by different degrees to achieve varied shapes. It can be the fence for pen holder or a stand for small items.}
    \label{fig:12}
    \end{figure*}

\subsection{Applicability to Traditional Woodworking Tasks}

This section focuses on traditional woodworking tasks and shows three demonstrations. 

\paragraph{Demo 1: Fabricating joinery.} When one of a pair of joints is available (the other pair is lost or broken), the user can utilize \systemName to create the corresponding half using a progressive method. We provide two examples: an edge joint (Figure~\ref{fig:10} (a-d)) and a T-joint (Figure~\ref{fig:10} (e-h)). The shape of the existing edge joint has two varied-length feet on each edge. The existing T-joint has a U-shaped pocket with two depths in the center of the board. We intend to use \systemName to create matched joints for them.

\paragraph{Demo 2: Furniture making.} This involves creating parts that interlock together. We demonstrated this by creating a torii. As in traditional carpentry, the user must arrange the layout accurately using a ruler. The wood pieces to be cut can be matched by measuring and marking their lengths (Figure~\ref{fig:11}(a)). Unlike traditional woodworking or CAD, which require the user to control the cut, \systemName users only need to draw the purple outline and red cross (Figure~\ref{fig:11}(b)), which are then set on the machine tool one at a time, and the CNC does the rest automatically (Figure~\ref{fig:11}(c)). Users can quickly complete inset work at the end (Figure~\ref{fig:11}(d)).

\paragraph{Demo 3: Kerf bending} It considers the cut depth, distance, and wood softness while creating modern curves and flexible shapes. We demonstrated how to create a ringed fence using a straight wood piece (Figure~\ref{fig:12}). In addition to this demonstration, there are other projects that combine kerf manufacturing with high-detail engraving, usually to improve aesthetics or produce useful items with fine details, such as artists and wearable art \cite{woodart}.

Through the above examples, we demonstrated that \systemName can produce a wide range of geometries to complete typical woodworking tasks, such as joinery, making an assembly from interlocking parts, and aesthetic engraving (kerf).

\subsection{Workshop}

\begin{figure*}[!htb]
\includegraphics[width=0.9\textwidth]{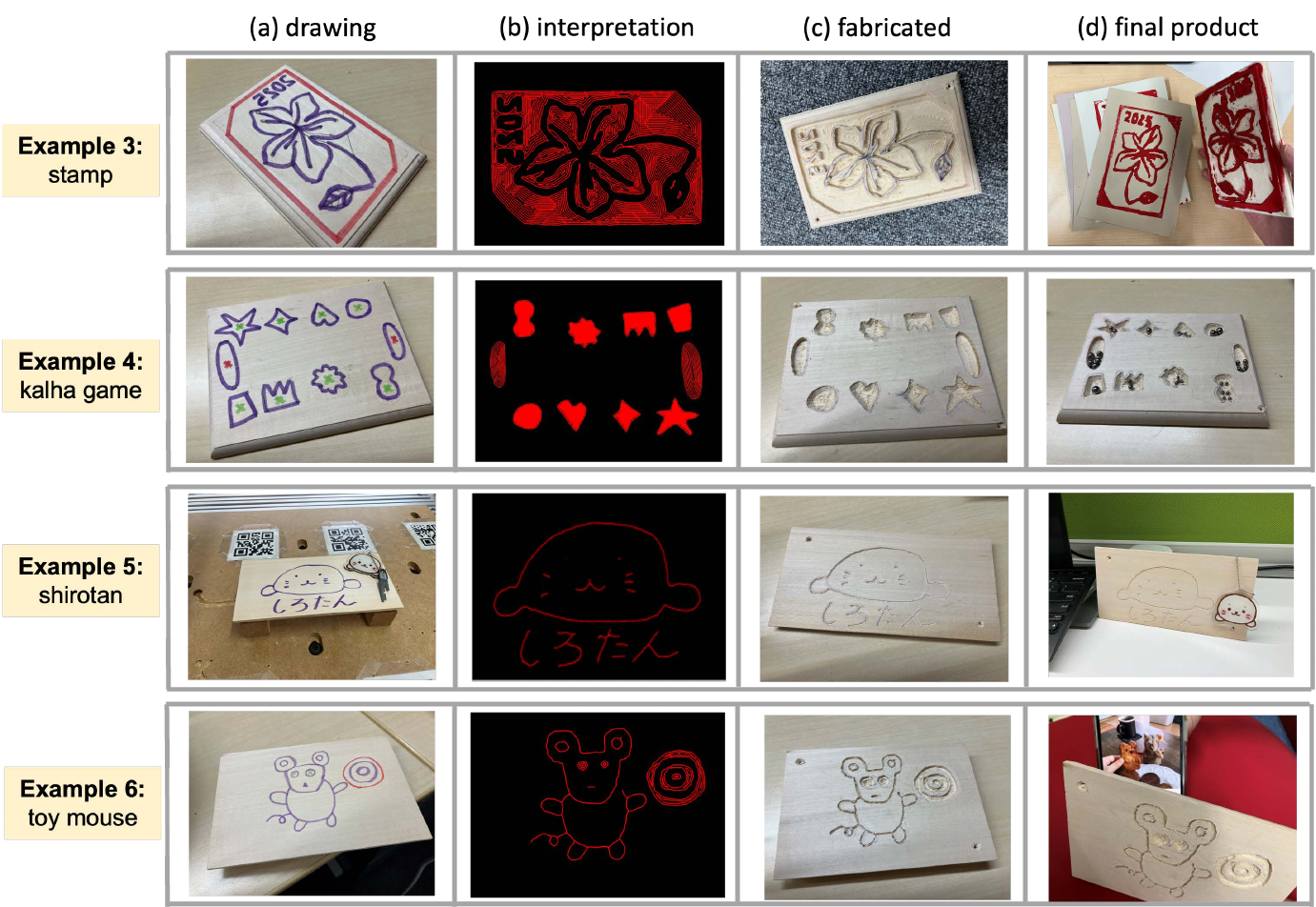}
\caption{Products designed and fabricated in the workshop: (a) user’s drawing results, (b) \systemName interpretation of drawing intent and visualization in red trajectories, (c) CNC-fabricated output; and (d) how users interact with the output.}
\Description{Products designed and fabricated in the workshop: (a) user’s drawing results; (b) \systemName interpretation of drawing intent and visualization in red trajectories; (c) CNC-fabricated output; and (d) how users interact with the output.}
\label{fig:13}
\end{figure*}

We run a workshop with five participant to understand how usable our system is for end users. We provided participants with a quick explanation of our language and asked them to draw on the wood to express their fabrication intent. The wood was then placed on a machine for complete fabrication. Finally, we provided them with the results and conducted a quick interview to gauge their satisfaction with the final output and whether the final combination was consistent with the design they intent (Figure~\ref{fig:13}).

\begin{figure*}[!htb]
\includegraphics[width=1\textwidth]{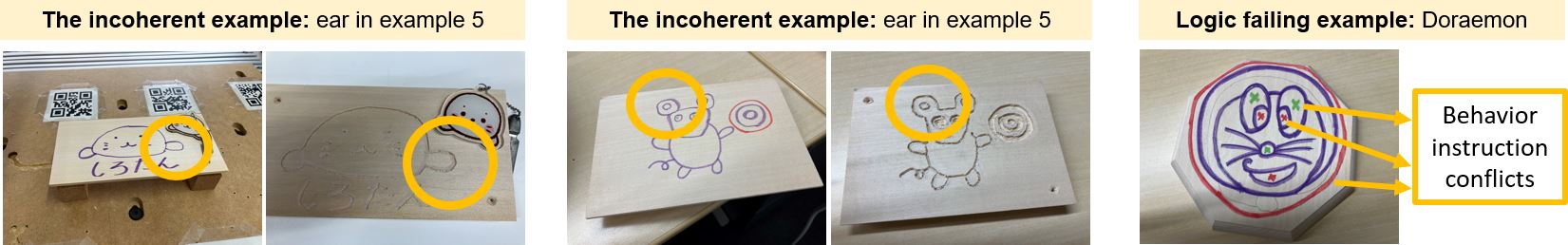}
\caption{The incoherent part and failure demo in the workshop.}
\Description{The incoherent part and failure demo in the workshop.}
\label{fig:14}
\end{figure*}

In our participant demonstrations, the two products completely fulfilled this goal. One participant (female, age approximately 30 years, an experienced wood joint designer) expressed her outlook for the coming year by designing a flower and the year 2025, intending to use it as a wood block print for the new year’s cards, as mirrored on the wood in Example 3. The second demo was created by a 7-year-old girl with no experience in fabrication. She revealed her love for the board game \textit{Kahla} (also known as \textit{Mancala}) and designed a one-of-a-kind \textit{Kahla} game by drawing it with specific pit forms (see Example 4). Two other demos were successfully generated; however, there were flaws in trajectory planning. Two participants (both male and graduate students with no experience in wood fabrication) created their favorite characters on wood. One is the anime \textit{Shirotan} (Example 5) and the other is \textit{Toy Mouse Holds a Doughnut} (Example 6). These four outcomes demonstrate that our approach enables users to design a functioning item that satisfies their particular needs with a minimal technical understanding of digital manufacturing. 

Drawing such cartoon characters involves joining numerous line segments together (Figure~\ref{fig:14}). When multiple strokes traverse the same point, the intended trajectory of the system skips certain difficult-to-pass line portions, similar to the \textit{Seven Bridges of Königsberg}. However, the participants did not notice this, and expressed pleasure with the final results. One participant sketched \textit{Doraemon} and experimented by combining several behavior pens (Figure~\ref{fig:14}). This result is inconsistent with the participants’ purpose, because both green and red represent cutting. When these two colors of behavior pens are combined in the same purple line, there is a disagreement. When red and green are grouped with the same purple contour or are nested, the system ignores the action represented by one of the colors.

\subsection{Demonstration of Advanced Parameter Control}

We observed that our participants enjoyed exploring intricate structures and shapes, and some liked controlling parameters other than shape or cutting type. That is, if we draw the same pattern in our language, the results may vary because of differences in settings. According to the participants’ feedback, we tried to adjust the three parameters by modifying the code and discussed the possible method to control them by user sketching in the future.

\begin{figure*}[!htb]
\includegraphics[width=1\textwidth]{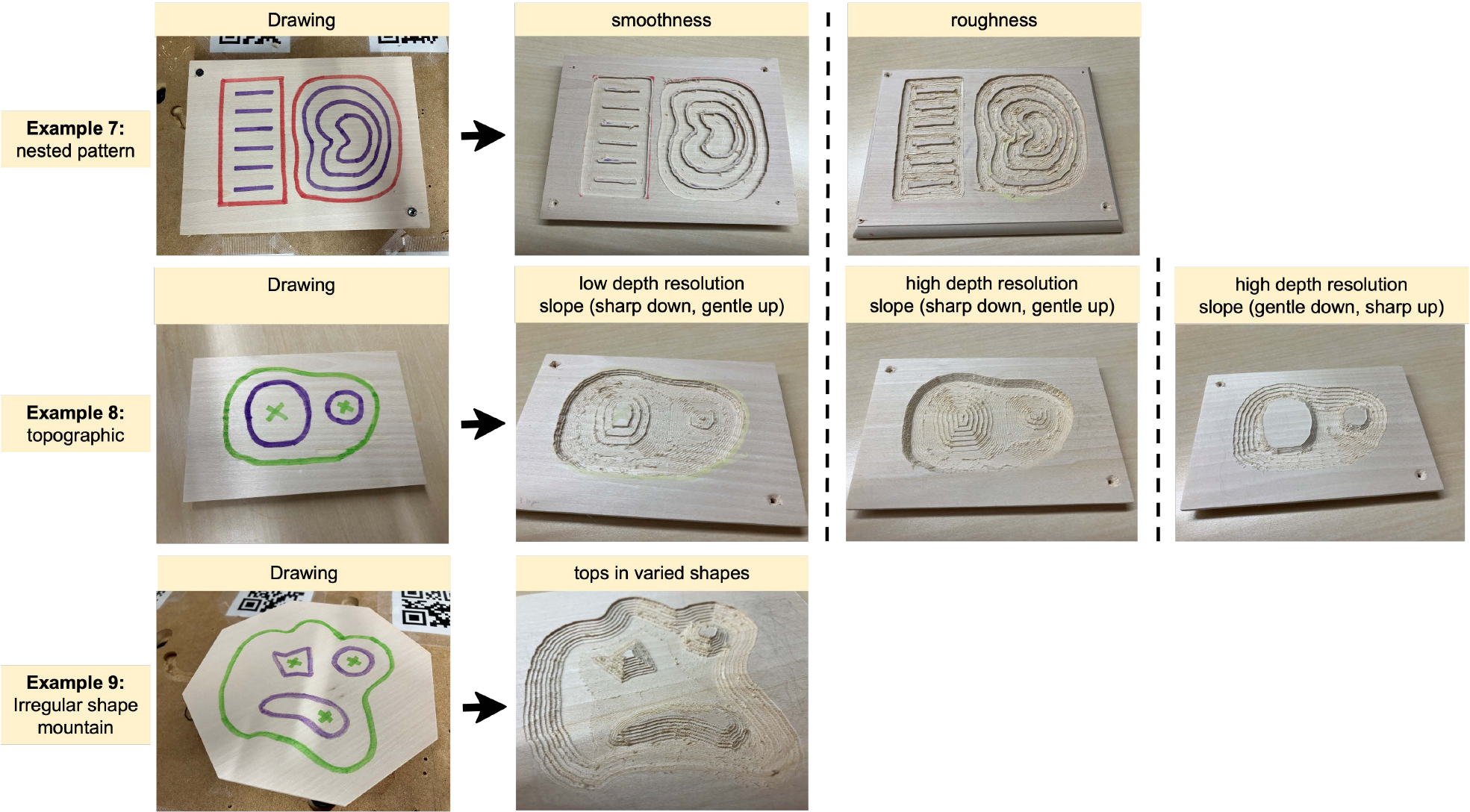}
\caption{Complexity in shape and parameters. We tested three parameters. Example 7 shows the surface smoothness or roughness. Example 8 shows the downward resolution and the steepness of the slope. Example 9 shows the flexibility of drawing shapes.}
\Description{Complexity in shape and parameters. We tested three parameters. Example 7 shows the surface smoothness or roughness. Example 8 shows the downward resolution and the steepness of the slope. Example 9 shows the flexibility of drawing shapes.}
\label{fig:15}
\end{figure*}

\begin{figure*}[!htb]
\includegraphics[width=0.9\textwidth]{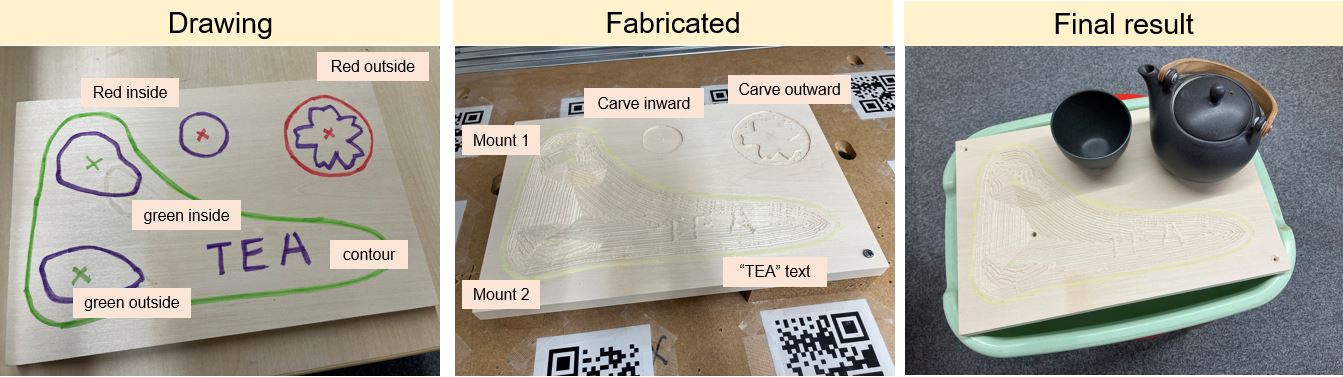}
\caption{Our capstone demonstration. The tea plate contains all language types we developed, with the combination of purple, red, and green functions.}
\Description{Our capstone demonstration. The tea plate contains all language types we developed, with the combination of purple, red, and green functions.}
\label{fig:16}
\end{figure*}

The first is plane refinement (smoothness or roughness), which refers to the distance the mill bit moves ahead with each step during cutting. We tested this in a nested structure drawing made using purple and red pens with random freeform shapes (Figure~\ref{fig:15}, first row). The greater the amount of refinement, the smoother the cut surface. However, some products require rough surfaces. In Example 7, the left part depicts a series of wooden blocks on the surface of the water, and the right part depicts a similar model of a Japanese garden, \textit{Karesansui}. When we reduce the fineness, our system can generate textured surfaces that resemble ripples on a lake. The remaining two parameters are depth resolution and slope (Figure~\ref{fig:15}, second row). We use a topographic contour line as an example. In Example 8, we drew two mountains and compared two degrees of depth resolution (low and high), as well as two slopes (one decreasing with a sharp slope and increasing with a gentle slope, and another switching the slopes). We propose that users have a high degree of freedom when drawing a line, including its relative position and shape (Figure~\ref{fig:15}, last row). Our system can assess the environment of the purple curves (their position in relation to the surrounding green markers) to design the shape of the mountaintop, such as a spire, flat top, or line (Example 9).

Based on a failed pilot study (\textit{Doraemon}), we discovered that users may employ both green and red pens to create intricate structures. We examined the boundaries of both behavior pens. We can simultaneously apply three different color strokes on the wood at the same time. In other words, users can combine behavioral marks applied both inside and outside a single purple shape to achieve a more complex effect. However, each purple contour must include one-on-one behavior directives (red or green). We demonstrated this by building a canyon-shaped tea table (Figure~~\ref{fig:16}, Example~10) and applying all types of languages. When \systemName is used, users do not need to worry about spatial measurements or technical difficulties. We propose that the system can allow designers to express themselves creatively during fabrication by eliminating the need to focus on technical details.

\section{Discussion}

\subsection{Participant Feedback}

When designing a drawing language, interaction with diverse products must be minimized to avoid complexity. In other words, we simplified users’ interactive operations while enabling them to produce interesting and diverse outcomes. Consequently, in addition to designing 2D languages (with purple and red pens), we investigated a limited number of 3D languages. A green pen was used to create curved surfaces or arches. Throughout the experiment, the following four constraints were identified: First, participants like to customize not only the shapes and cutting methods of the behavior pen, but also other characteristics such as the fineness (i.e., resolution) and slope. Therefore, in the future, we intend to allow users to modify these parameters using stroke attributes such as stroke thickness or color depth. For example, the thinner the stroke, the steeper the slope. The deeper the color (e.g., deep green or light green), the smoother the surface. Second, the users used a purple pen to create patterns with interlaced lines. Additionally, we intend to develop an algorithm that can produce grid-like patterns more efficiently. Third, we observed cases in which users drew multiple-colored strokes that overlapped or joined, and errors are likely to occur when the system recognizes the colors. Hence, we also need to improve the color-recognition system by considering color-mixing calculations. Fourth, these two behavior pens can be used together only in limited contexts.

\subsection{Comparison between \systemName, CAD, and Traditional Fabrication}

\recheck{Both CAD and \systemName can lower the typical fabrication barriers for novice and experienced users in different aspects. \systemName excels in offering high freedom and customization in manufacturing. Each artifact produced using \systemName is unique, reflecting the designer’s individual style. By leveraging users’ sketching skills, \systemName lowers the entry barrier for personal fabrication, enabling even children to design and manufacture toys. In contrast, manual woodworking requires specialized expertise, and CAD users need both hardware and programming skills. CAD offers clear advantages in handling complex designs. Its interface enables users to solve geometric problems with constraint-based relationships and design intricate interrelated components using timeline-based modeling. Additionally, CAD supports importing external libraries, such as joint libraries \cite{tian_matchsticks_2018}, perspective tools \cite{larsson_tsugite_2020}, or material composition analysis for parameter setting \cite{dogan_sensicut_2021}.}

\recheck{While some tasks can be accomplished using either CAD or \systemName, the difficulty depends on the type of task. For instance, designing regular shapes is easier with CAD but harder with human hands, while freeform designs are more intuitive for humans and difficult to handle with CAD. Similarly, in drill pattern fabrication, where a series of precisely spaced holes are created based on a predefined layout, humans must manually mark each point, while CAD streamlines the process with a single click. However, tasks involving varied settings or sequence regulation are easier with \systemName, as users can simply apply differently colored pens, while CAD requires creating separate toolpaths. To objectively compare CAD and \systemName, aspects such as fabrication accuracy, aesthetic expression, and challenges in complex geometric manufacturing must be analyzed. This can be achieved through user studies and interviews with woodworkers to gather anecdotal data and conducting in-depth analysis, which is discussed in the following section.}

\section{Limitations and Future Work}

\subsection{Expanding to Multi-Axis CNC}

If a 4-axis CNC router is used, full 3D fabrication is possible. The 4th axis milling typically refers to cutting processes that use a rotary table. There are two possible ways to achieve this goal. The first is the addition of more colors. However, multicolor designs become increasingly difficult as the number of combinations increases exponentially. In addition to providing more colors, we investigated how users could draw on both the top and sides of the physical material. To realize this, users need to plan considering different perspectives. However, some users, particularly children, find this difficult to achieve. Developing a language that allows users to draw instinctively rather than mastering the skill of drawing from a perspective is a potential direction for future investigations. In addition, interlocking, mortise, and tenon structures are difficult to produce with CNC machines, even if they can be drawn on both sides of the wood. The solution we developed involves adding audio-based feedback to tell the user how to flip the wood to start cutting on the other side after the CNC machine has completed one side of the cut.

\subsection{Hardware Limitations}

We tried several methods to increase the accuracy and collected multi-frame data to filter out noise when creating a virtual workspace. Additionally, we discovered that the offset was stable; therefore, we manually adjusted for it. In the future, two methods can be used to increase the accuracy. First, we may test the system using a high-resolution depth camera. Second, numerous depth cameras may be positioned at various locations throughout the workspace. RGB cameras can be complemented by NIR cameras to detect hidden QR codes for more unobtrusive surface registration~\cite{dogan_infraredtags_2022, dogan_brightmarker_2023}. Furthermore, CNC have an infinite number of end mills for various cuts. Each milling bit required its own feed rate and spindle speed. Throughout the development and experimentation, we continued to use one type (tapered ball nose carving end mill 1/16 in), which was narrow and had a slight slope. This mill bit is suitable for 3D carving \cite{ballnose} but not for an assembly pocket. Because the upper half was thicker than the lower half, the sliced shape expanded when cut to a specific depth. In the future, we will generalize this language to cover multiple forms of mill bits.

\subsection{Generalizability to Other Subtractive Manufacturing Processes}

First, we examine the distinctions between other machines. Laser and waterjet cutting typically cut through materials that cannot be used for 3D carving jobs~\cite{dogan_structcode_2023}. If the existing \systemName language is used directly, only purple and red pens are required. The properties of the laser cutter are investigated in detail. When different powers are used, they can cut through materials (such as metal) and heat specific lines of plastic sheets, allowing them to be folded to perform origami activities~\cite{kusunoki2024integration}. Consequently, new colors may be used to describe the power and speed of the laser cutter. We investigated the commercial use of waterjet and discovered that it has high requirements for the intricacy of 2D designs. Therefore, we should consider using different colors to represent the dashed cutting line or drill pattern. In addition to the current auto-smoothing function, the visualization interface requires a regularization function.

\subsection{User Study for Aesthetic Design}

We envision future work can more deeply investigate how our system improves design aesthetics. A follow-up research question of interest might be, “are the designs produced using sketch-based methods (e.g., \systemName) qualitatively different from those produced with a purely digital tool?”
As a follow-up study, we plan to invite participants of various backgrounds (experienced CAD users, artists, conventional fabricators, etc.) and ages (children, students, and working professionals), and provide brief instructions to create a design with different difficulty levels using \systemName and a digital tool (\textit{VCarve Pro}) independently. Finally, to address the research question, we plan to evaluate the interviews and compare the designs generated by a standard digital process to \systemName drawings. Our hypothesis is that the proposed system can enable designers to express themselves creatively and realize their full artistic potential.

\subsection{Expanding Personal Fabrication}

The proposed method lowering the entry barriers for subtractive personal manufacturing. However, the availability of the CNC machine itself is limited and therefore this tool cannot be categorized as a personal fabrication tool, even though we solve the technical issue. In the future, we hope to collaborate with wood workshops to commercialize the use of \systemName. Users can purchase wood, sketch the content at home using the \systemName language, take the content to the wood workshop, and place it directly on the \systemName-enabled CNC machine to achieve the desired output. Because not every family has a large printer, users occasionally transfer files to a USB and bring them to the printing shops. We hope that people will use the same workflow to get the fabricated woodwork of their own design. As for the mapping language, we plan to enable the language to become more adaptable so that users can set the colors themselves for corresponding functions.

\section{Conclusion}

In this study, we propose \systemName, a system that allows users to interact directly with materials-to-be-cut using physical instructions, thereby eliminating time-consuming measurement tasks and digital representation difficulties of the CAM process. While evaluating the design criteria, we defined both design requirements (language and visual interface) and technical requirements (registration). We conducted a technical evaluation to demonstrate the error distribution of the virtual-physical space alignment in our system and four-step experiments to show the advantages and general ability of \systemName. \systemName shows a new and promising direction for making the digital fabrication process more intuitive. This work has the potential to inspire future studies on how drawing instructions for fabrication can boost user creativity and engagement, and further development of the drawing language to support the creation of even more complex designs.

\begin{acks}
This work was supported by JST ACT-X Grant Number JPMJAX210P, JSPS KAKENHI Grant Number JP23K19994, JST AdCORP Grant Number JPMJKB2302 and a collaborative research fund between Mercari Inc. R4D and RIISE, The University of Tokyo.
\end{acks}

\bibliographystyle{ACM-Reference-Format}
\bibliography{drawn2cut}


\end{document}